\documentclass{jfm}

\usepackage{graphicx}
\usepackage{subcaption}
\usepackage{epstopdf,epsfig}
\usepackage{natbib}
\usepackage{rotating}
\usepackage{color}
\usepackage{xcolor}
\usepackage{float}
\usepackage{amsmath}
\usepackage{verbatim}
\usepackage{soul}

\newcommand{\rd}{\mbox{d}}
\newcommand{\rt}{\tilde{r}}
\newcommand{\nt}{\tilde{\nabla}}
\newcommand{\rJ}{\rm J}
\newcommand{\rI}{\rm I}

\let \d=\partial

\let\qq=\qquad
\let\vp=\varphi

\let\vp=\varphi

\shorttitle{Hydro-elastic effects during the fast lifting of a disc from a water surface}
\shortauthor{\qquad P. Vega-Mart\'{\i}nez, J. Rodr\'{\i}guez-Rodr\'{\i}guez,

T. I. Khabakhpasheva and A. A. Korobkin}

\title{Hydro-elastic effects during the fast lifting of a disc from a water surface}

\author{P. Vega-Mart\'{i}nez\aff{1}
  \corresp{\email{pvega@ing.uc3m.es}},
  J. Rodr\'{i}guez-Rodr\'{i}guez\aff{1},
  T. I. Khabakhpasheva\aff{2} 
 \and A. A. Korobkin\aff{2}}

\affiliation{\aff{1}Fluid Mechanics Group, Universidad Carlos III de Madrid, 28911 Legan\'es, Madrid, Spain
\aff{2}School of Mathematics, University of East Anglia, Norwich NR4 7TJ, United Kingdom}

\begin{document}

\maketitle

\begin{abstract}
Here we report the results of an experimental study where we measure the hydrodynamic force acting on a plate which is lifted from a water surface, suddently starting to move upwards with an acceleration much larger than gravity.
Our work focuses on the early stage of the plate motion, when the hydrodynamic suction forces due to the liquid inertia are the most relevant ones. Besides the force, we measure as well the acceleration at the center of the plate and the time evolution of the wetted area. The results of this study show that, at very early stages, the hydrodynamic force can be estimated by a simple extension to the linear exit theory by \citet{Korobkin2013}, which incorporates an added mass to the body dynamics. However, at longer times, the measured acceleration decays even though the applied external force continues to increase. Moreover, high-speed recordings of the disc displacement and the radius of the wetted area reveal that the latter does not change before the disc acceleration reaches its maximum value. We show in this paper that these phenomena are caused by the elastic deflection of the disc during the initial transient stage of water exit. We present a linearised model of water exit that accounts for the elastic behaviour of the lifted body. The results obtained with this new model agree fairly well with the experimental results.
\end{abstract}


\section{Introduction}

Marine and naval structures commonly experience impact loads caused either by waves or by the motion of the structure into and out of water. For instance, ship sections enter the water and then subsequently exit it in rough sea conditions, a process known as slamming. A related phenomenon, wetdeck slamming, is observed in offshore engineering, where a wave hits the wetdeck of a platform from below, a problem studied both experimentally and numerically by \cite{Baarholm2001}, \cite{Baarholm2004}, \cite{Faltinsen2004}, and \cite{Scolan2006}. During a first stage, the wetted area of the deck increases in time leading to high hydrodynamic loads. Then, during the second stage, the wetted area diminishes as the water falls down due to gravity.

It is common wisdom that a body that enters a water surface at high speed may experience very large forces. However, it is less obvious that forces of similar magnitude --but opposite sign-- act in the opposite case, i.e. when the body exits water. Commonly,  hydrodynamic loads are defined as positive during the entry stage, and negative during the exit one. Here, negative loads mean that the hydrodynamic pressure over the wetted part of the ship hull is below the atmospheric pressure, albeit still above the vapour pressure at which cavitation occurs. It is remarkable that, although the exit stage lasts longer than the entry one, the magnitude of the negative loads can be comparable to the magnitude of the positive ones \citep{Faltinsen2004, Scolan2006, PiroMaki2011}. Despite these similarities, the physical origins of these loads are different. While the positive loads during the entry stage scale with the dynamic pressure and thus are proportional to the entry velocity of the body squared, negative loads during the exit stage are governed mainly by the acceleration, provided this is large. Conversely, if the body leaves the water slowly, then gravity and hydrostatic pressure play the major roles \citep{Greenhow1988, rajavaheinthan2015}. Despite water entry being much related to water exit, while the former problem has been studied extensively in the past \citep{KorPuk1988}, the latter has received comparatively less attention.

We mention here several studies that emphasize the water exit stage. Two-dimensional problems of water entry and exit were investigated numerically by \cite{PiroMaki2011, Piro2012, Piro2013} using full Navier--Stokes simulations. In particular, they computed the impact of a rigid wedge with deadrise angle of 10$^\circ$, initial velocity 4 m s$^{-1}$ and a deceleration of 92 m s$^{-2}$. They found that the hydrodynamic force is initially positive and then becomes negative even though the wedge continues to penetrate the water. Later, the magnitude of the negative hydrodynamic force reaches a maximum at the end of the entry stage, when the speed of the wedge is zero, and continues to be negative with a decreasing magnitude during the exit stage.

Recently, the development of analytical and semi-analytical approaches to calculate hydrodynamic loads in water-exit problems has become of interest. \cite{Korobkin2013} developed a linearised model that was in excellent agreement with existing computational results \citep{PiroMaki2011}. In this model, both the hydrodynamic equations and the boundary conditions were linearised by exploiting the fact that, at small times,  displacements are small. In particular, this allows us to impose the boundary conditions on both the free surface and the surface of the moving body on the initial equilibrium level of the water. The actual shape of the body was not included in the model, as well as gravity, surface tension and viscosity. The problem was then formulated for an acceleration potential, which is zero on the free surface.
The wetted part of the body surface diminishes during water exit. The speed of the contact line, which bounds the wetted part of the body surface, is set to be proportional to the local velocity of the flow along the body surface. This condition was introduced by \cite{Baarholm2004} and \cite{Faltinsen2004} in a two-dimensional problem of wetdeck slamming with the coefficient of proportionality between fluid and contact line speeds, $\gamma$, being equal to one. This is, the contact line was assumed to be a material one. \cite{Korobkin2013} set $\gamma=2$ using the computational results by \cite{PiroMaki2011} for the hydrodynamic force during the exit stage. The linearised model of water exit with a constant deceleration provided hydrodynamic forces which are very close to those computed with the Navier--Stokes solver (see figure 5 for a wedge and figure 6 for a parabolic contour in \cite{Korobkin2013}). The comparison becomes even better if nonlinear effects and gravity are included in the exit model by using the approximations suggested by \cite{korobkin2004} and \cite{khabakhpasheva2016}.

The linearised model of two-dimensional water exit was generalised, and validated using CFD results, by  \cite{Korobkin2017} to include arbitrary motions of bodies and bodies of time-varying shapes. The model was developed further to be included in the two-dimensions-plus-time analysis of aircraft ditching. \cite{KorKhaba2017} presented another model of water exit, which is based on a small-time asymptotic solution of the two-dimensional problem of a plate lifted suddenly from the water surface and the method of matched asymptotic expansions. A similar model was developed by \cite{iafrati2008} for water impact problems. In this theoretical study of water exit the flow near the edges of the plate and the motion of the contact line are nonlinear and self-similar, in contrast to the flow in the main region, which is linear at leading order for small displacements of the plate. The acceleration of the contact line was found to be proportional to $t^{-\frac{2}{3}}$ for a constant acceleration of the plate lifting, where $t$ is the time. The predicted shape of the free surface near the contact points was found to be very close to full Navier--Stokes numerical simulations. Note that, in the linearised exit model, the free-surface shape is very different from the computed one, despite the fact that the theoretical hydrodynamic forces compare well with experiments.

A simplified description of water entry and exit is provided by the von-K\'arm\'an model \citep{vonKarman1929}, which defines the wetted part of a body surface during both the entry and exit stages as the intersection of the body surface with the flow region determined without the effect of the body presence. This idea was used in water exit problems by \cite{Tassin2013} to estimate the hydrodynamic loads for bodies with time-dependent shapes. However \cite{Faltinsen2004} pointed that, using the von-K\'arm\'an approach, the calculated duration of the water-exit phase of a body is shorter than the experimentally observed one. Despite this discrepancy in the calculated duration of these stages, the peak values of the forces obtained with this method were still comparable in magnitude. In our problem, this model cannot be used because the flat disc, which has negligible draft, leaves the initial liquid region immediately after it starts to move upwards.

Due to the complexity of the exit flows with unsteady free surfaces, both numerical and theoretical findings and models should be validated against experiments. To the best of our knowledge there are few works that can be used to validate experimentally the kinematics of water-exit flows, but not the hydrodynamic loads, which are difficult to measure. However, two papers are worth highlighting as they focus specifically on water-exit \citep{Reis2010, Tassin2017}. The seminal study by \cite{Reis2010} deals with the volume of water dragged by a lifting disc that mimics a cat's tongue. This disc was controlled by a computer to reproduce the observed motion of an actual tongue, in particular reaching accelerations as high as 27 m s$^{-2}$. The dragged water volume was determined using high-speed imaging. They concluded that the domestic cat laps exploiting ``fluid inertia to defeat gravity and pull liquid into the mouth". Despite the interest of the experiment, these authors did not measure the forces required to pull the disc, so their results cannot be used to validate water exit theories \citep{Korobkin2013, KorMaki2013}. \cite{Tassin2017} reported experiments where they measured the dynamics of the contact line, or in other words the evolution of the wetted area. The acceleration of a circular transparent plate was measured to be between 0 and 25 m s$^{-2}$. Furthermore, they showed that the radius of the wetted region closely follows the radius predicted by the linearised exit model with $\gamma=1$. Recently, see \cite{tassin2018}, experiments with a cone lifted from the water surface were announced but the results have not been published yet. Again, in these studies the forces and hydrodynamic pressures during water exit were not studied.

Some ideas about the hydrodynamic forces during water exit can be gained from experiments with a body moved by a given external force. Then the measured acceleration of the body motion can be converted to the evolution of the hydrodynamic force acting on the body. This idea was employed by \cite{Korobkin_etal17CTAM2014} who studied experimentally, numerically and theoretically the oscillations of a rigid sphere entering and exiting a water surface. The sphere was supported by a spring with the equilibrium position above the water surface.
These experiments demonstrated that hydrodynamic force was negligible in the conditions reported. Theoretical analyses revealed that, for these forces to be relevant, the added mass of the wetted sphere should be at least comparable to the body mass. Moreover, it was concluded that the hydrodynamic force would be more visible if the body moves at a higher acceleration. These conclusions follow from the linearised exit model, which states that the hydrodynamic force is proportional to both the added mass of the body and the body acceleration.

With the aim of quantifying experimentally the hydrodynamic loads during water exit, in this work we present an experimental investigation of the loads acting on a circular plate, a disc, lifted impulsively from a water surface with a high acceleration.
Initially the disc touches the flat water surface with a negligible draft. Then the disc is moved suddenly upwards by a given external force applied at its center. Both the pulling force and the disc acceleration are measured with high temporal resolution in order to resolve the initial few milliseconds of the motion, when the highest loads occur. Besides force and acceleration, we use high-speed cameras to record the disc displacement and the radius of the wetted area. The disc acceleration peaks at about 200 m s$^{-2}$ in our experiments and then rapidly decays despite the fact that the external force continues to increase. We also observe that the wetted area of the disc does not change before the disc acceleration peaks. This behaviour of the disc acceleration and its wetted area is attributed in this study to the interaction between the hydrodynamic loads and the elastic deflection of the disc.

To explain and quantify theoretically these measurements, we introduce a model that solves the unsteady axisymmetric flow generated by a lifted elastic body coupled with its elastic deformations. This new model explains both the non-monotonic relation between the disc acceleration and the applied external force and the delay in shrinking the wetted part of the disc surface. The model is finally validated quantitatively by dedicated experiments with circular discs of different rigidities.

The paper is structured as follows: the experimental set-up and some results that justify the hypotheses adopted thereafter are presented in section 2. In section 3, the relation between the measured force and acceleration is compared to that predicted by the linearised theory of water exit of \cite{Korobkin2013}. Furthermore, the motion of the contact line is used to validate some results reported in \cite{KorKhaba2017}. In section 4, we present a generalised linear theory of water exit that takes into account elastic effects in the motion of the plate. Finally, some conclusions are presented in section 5.

\section{Experimental study of the water exit }\label{sec:experiments}

Here, we present the facility designed to measure the hydrodynamic forces acting on an object suddenly lifted from a water surface at an acceleration much larger than gravity, $a \gg g$. Besides the force pulling the object and its acceleration, we also measure the evolution of the wetted area or, conversely, the recoil of the liquid-gas-solid contact line.

\begin{figure}
\centering
\begin{tabular}{rl}
\raisebox{50mm}{$a)$} & \includegraphics[width=0.9\textwidth]{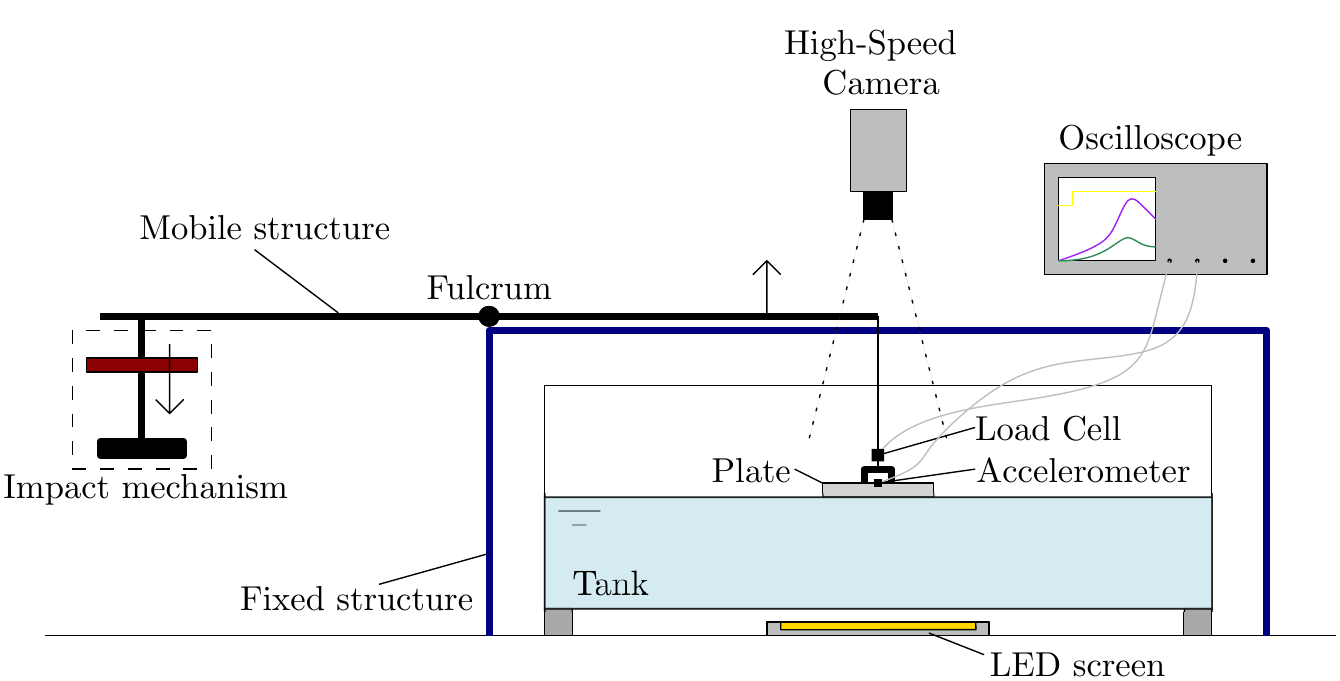}\\
\raisebox{50mm}{$b)$} &
\includegraphics[scale=0.9]{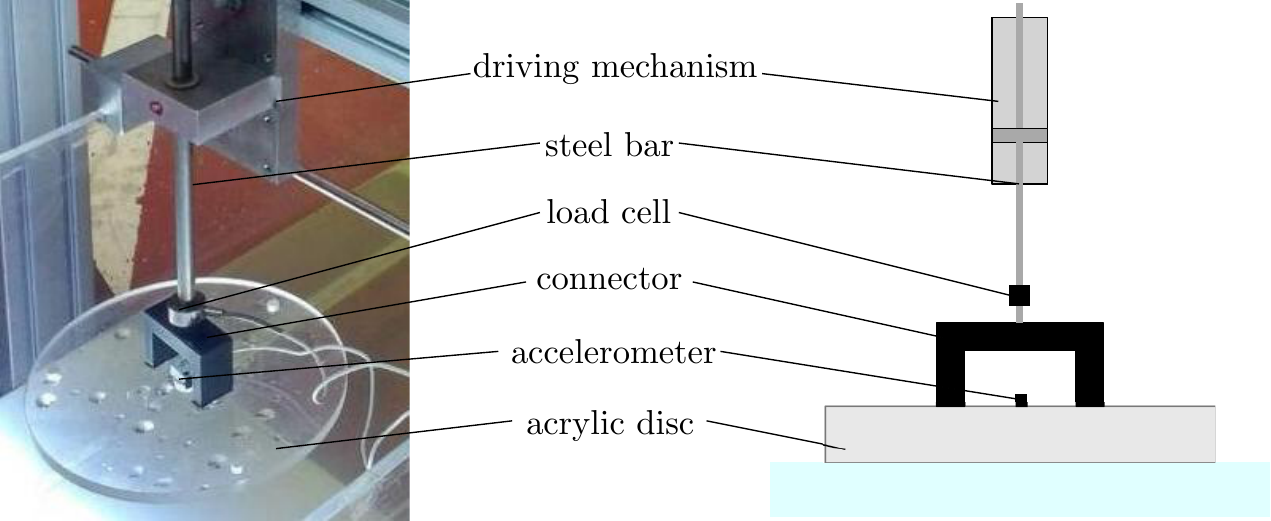}
\end{tabular}
\caption{\label{fig:experimental_setup} $(a)$ Layout of the experiment. $(b)$ Detail of the plate, which is connected to a driving mechanism through an inverted U-shape structure fixed at its upper surface.}
\end{figure}

The layout of the experimental apparatus is shown in figure \ref{fig:experimental_setup}.
The plate is set in motion by a structure which operates as a catapult. A mobile arm, a seesaw, turns around a fulcrum located in a fixed structure anchored to the ground. The plate is supported by a steel bar that hangs from one end of the mobile arm through a steel cable. An auxiliary structure guides the movement of steel bar, to guarantee that its motion is purely vertical. The other end of the arm is connected to the impact mechanism, which consists in a weight that slides along a rail and a base that stops its fall and thus transmits the impact to the seesaw. 

\begin{figure}
\centering
\begin{tabular}{lr}
\includegraphics[width=0.5\textwidth]{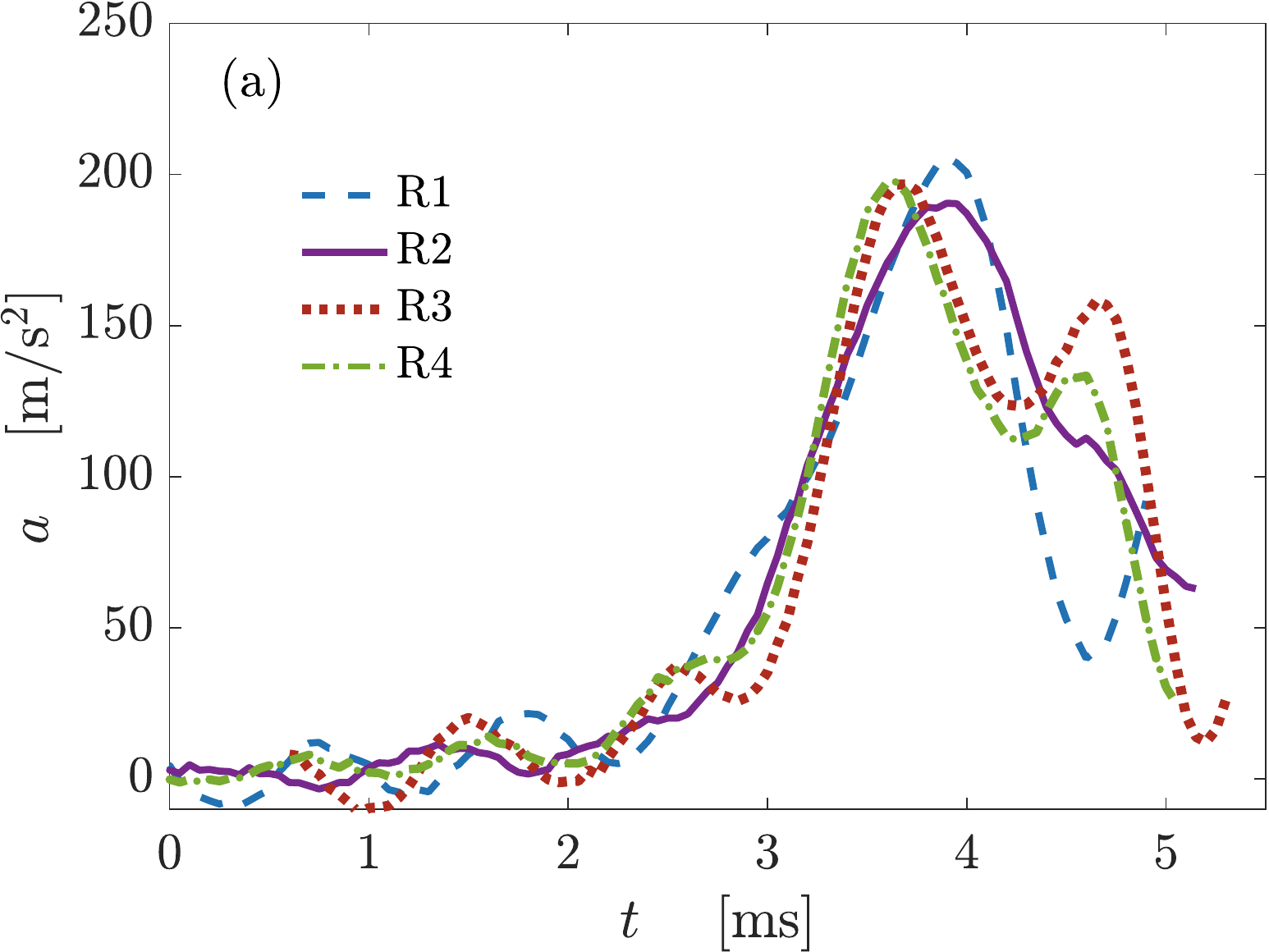} &
\includegraphics[width=0.5\textwidth]{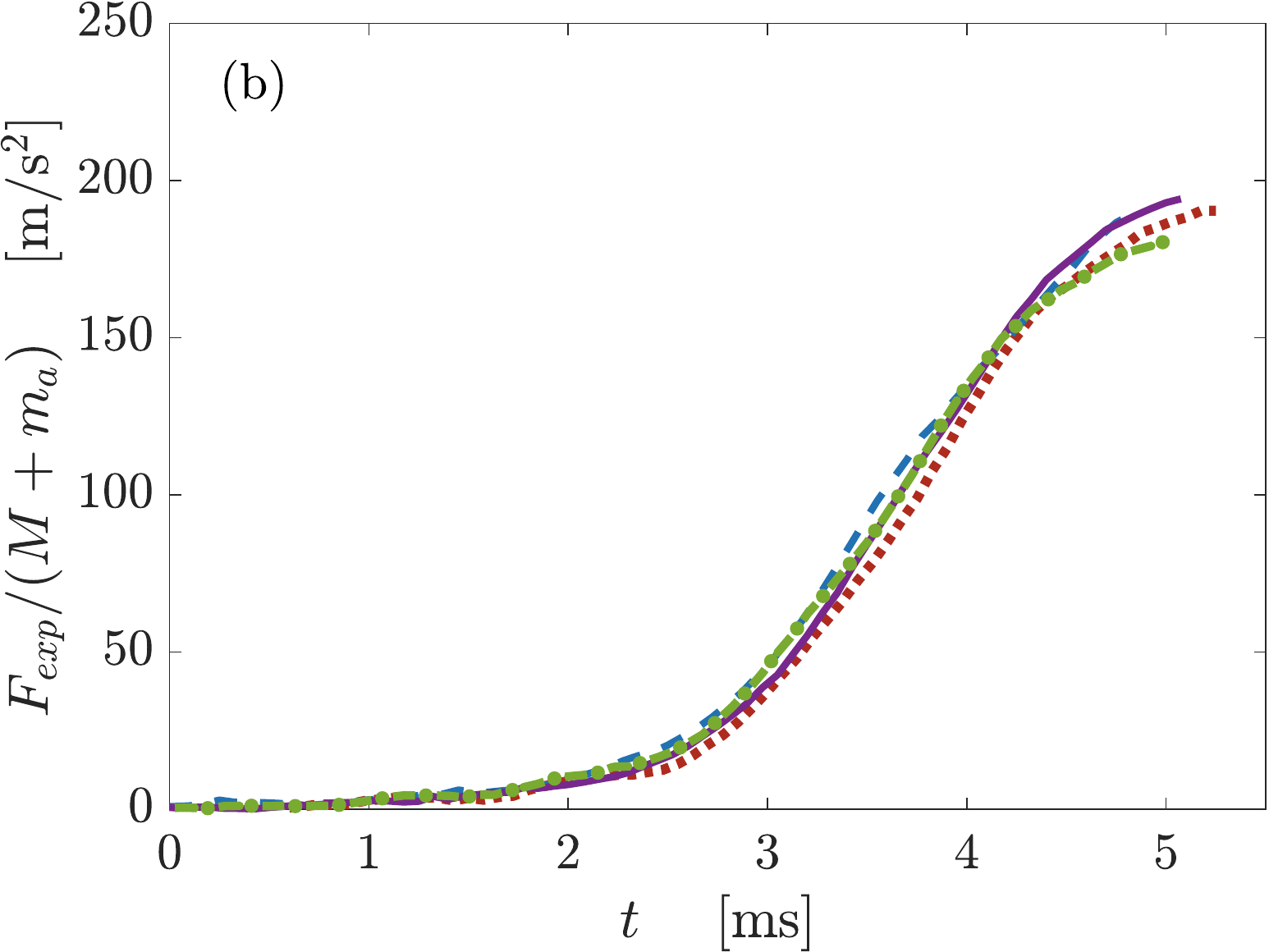}
\end{tabular}
\caption{\label{fig:repeatability}Comparison between acceleration and force measurements obtained under identical experimental conditions for four repetitions of session 3.}
\end{figure}

The bar connects to the plate through a fast-speed-of-response load cell (Honeywell Model 31 mid, 100 lb), capable of measuring forces of up to 450 N. The disc is equipped with an accelerometer (Honeywell model JTF $\pm$ 50G), placed at its centre. The signals from the load cell and the accelerometer are read by an oscilloscope (Tektronik TDS3014c, sample rate of 250 kHz), after being pre-conditioned by amplifiers (ADAM 3016). During postprocessing, the acceleration of gravity and the weight of the plate are subtracted from the results. To illustrate the good repeatability of the results, we plot in figure \ref{fig:repeatability} the acceleration and force corresponding to wet experiments 
performed under identical conditions (see table \ref{tab:sessions} for details). To make clear the point that the curves are essentially the same, within the experimental variability, they have been shifted a small time to compensate for the uncertainty in the determination of the impact time ($\Delta t_1 =$ 0.2 ms, $\Delta t_2 =$ 0.7 ms, $\Delta t_3 =$ -0.6 ms and $\Delta t_4 =$ 0.85 ms).

The plate, which properties are summarised in table \ref{tab:properties}, is made of transparent acrylic to allow for the observation of the contact line motion. The total mass of the equipped disc, $M$, is the sum of the mass of the plate, $M_p$, the connector, accelerometer and the load cell, $M_c$.  The connector and the steel bar between the load cell and the driving mechanism are considered as perfectly rigid in the present analysis. They are made up of aluminium, much stiffer than the material of the plate.

\begin{table}
   \centering
    \begin{tabular}{c c}
    
     \hline
     Radius, $R$ &  10.8 cm \\
     Thickness, $h_p$ & 1 cm \\
     Mass of the plate, $M_p$ & 0.432 kg\\
     Mass of the equipment, $M_c$ & 0.198 kg\\
     Added mass of the plate, $m_a$ & 1.680 kg \\
     Total mass, $M=M_p+M_c$ & 0.630 kg \\
     Density, $\rho_p$  & 1180 kg m$^{-3}$ \\
     Young's modulus, $E$ & $3.1\times 10^9$ Pa \\
     Poisson ratio, $\nu$ & 0.33 \\
   \hline
    \end{tabular}
    \caption{Summary of the features and material properties of the plate.}
    \label{tab:properties}
\end{table}

At the beginning of the experiment, the weight (5 kg) is released and allowed to fall freely (10 cm), sliding down the rail until it impacts the base. When this happens, this instant is regarded as the time origin, $t = 0$.

We performed two kind of experimental sessions, some where the plate is touching the water surface of the tank (wet experiment), and another one, identical in everything to the first one, except in that the disc does not touch the water surface (dry experiment). Wet experiments are carried out with the disc touching the water surface of a tank, which dimensions, 100 x 40 x 40 cm, are large enough to guarantee negligible boundary effects.

\begin{table}
    \centering
    \begin{tabular}{cccc}
    Session & Dry/Wet & $M$ (kg) & $h_p$ (cm) \\ \hline
    1 & Dry & 0.630 & 1 \\
    2 & Wet & 0.630 & 1 \\
    3 & Wet & 0.630 & 1 \\
    4 & Wet & 1.070 & 2 \\
    5 & Wet & 0.430 & 0.5 \\ \hline
    \end{tabular}
    \caption{Summary of the experimental sessions. Sessions 2 and 3 are identical in their conditions, but high-speed movies were acquired only for session 3.}
    \label{tab:sessions}
\end{table}

\begin{figure}
\centering
\begin{subfigure}[b]{0.23\textwidth}
\includegraphics[width =\textwidth]{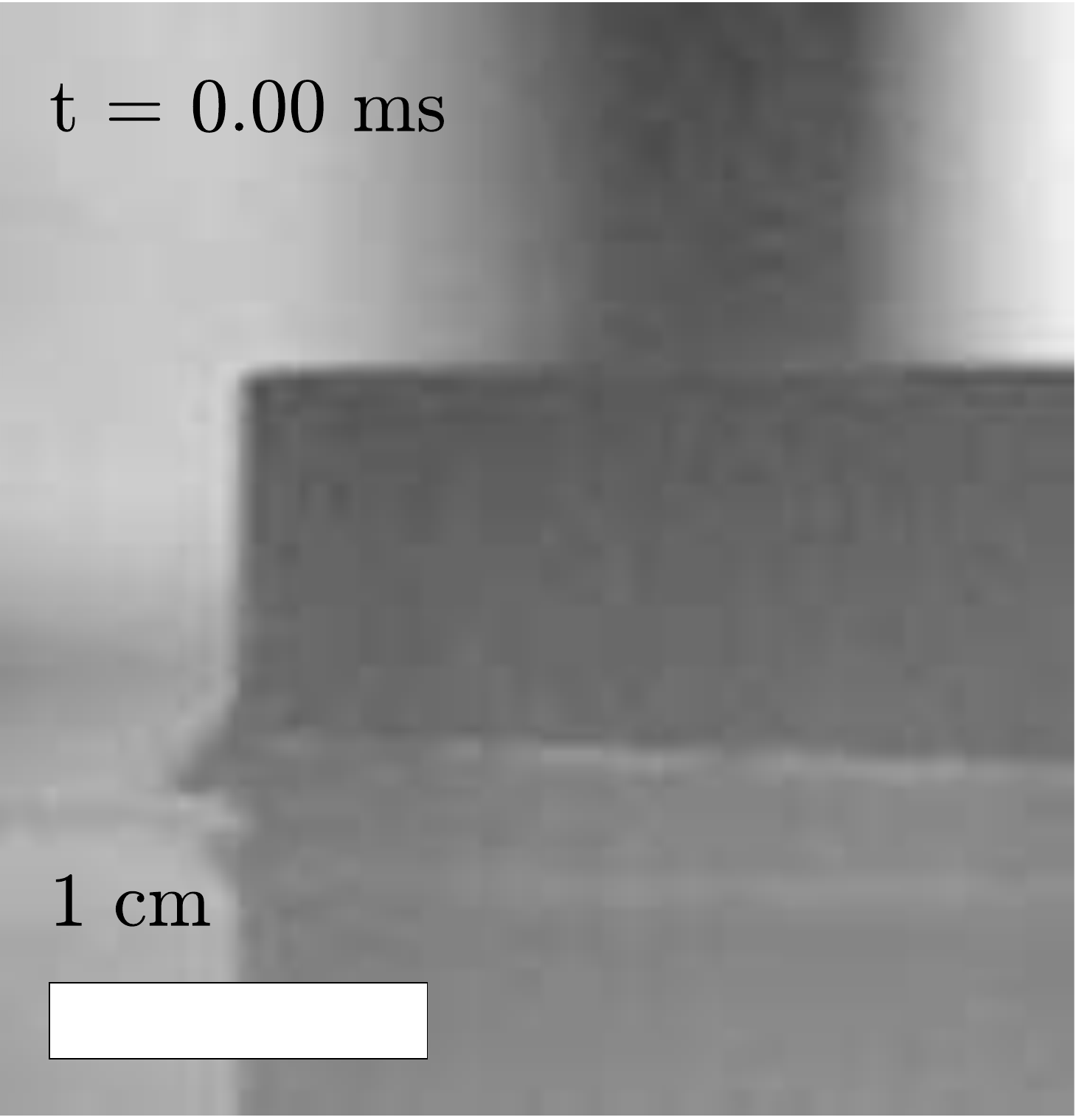}
\caption{\label{fig:1}}
\end{subfigure}
\begin{subfigure}[b]{0.23\textwidth}
\includegraphics[width =\textwidth]{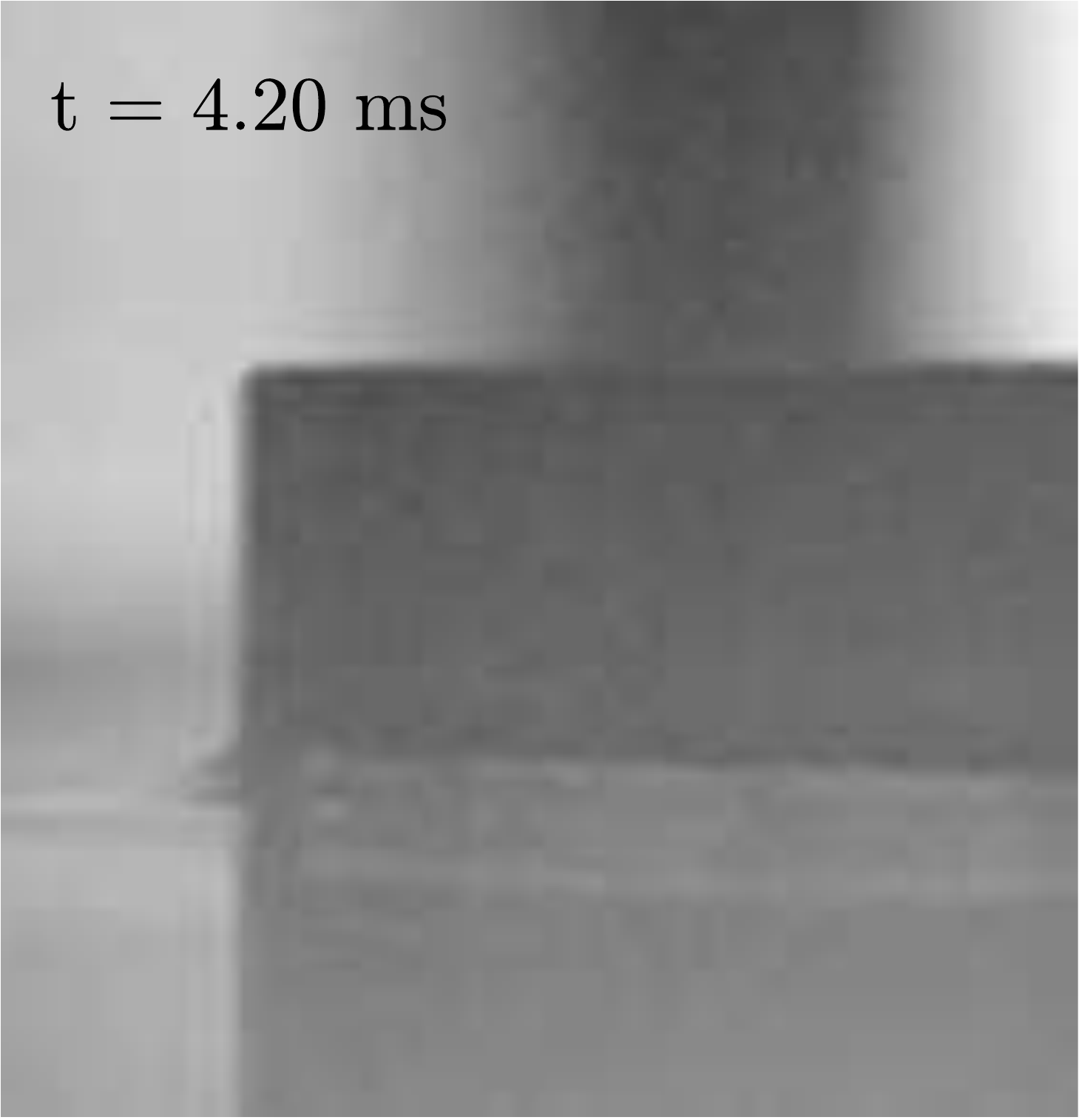}
\caption{\label{fig:2}}
\end{subfigure}
\begin{subfigure}[b]{0.23\textwidth}
\includegraphics[width =\textwidth]{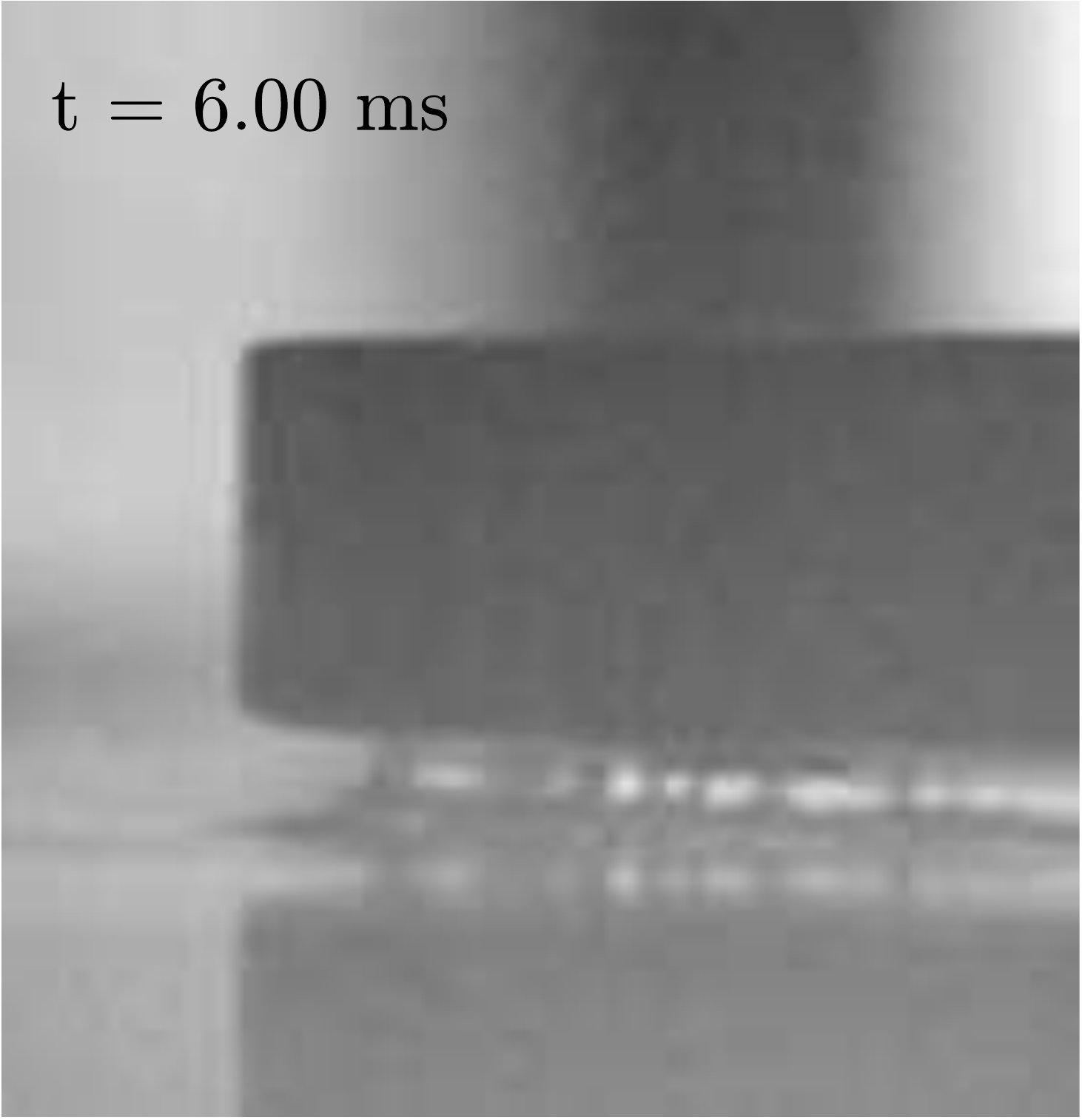}
\caption{\label{fig:3}}
\end{subfigure}
\begin{subfigure}[b]{0.23\textwidth}
\includegraphics[width =\textwidth]{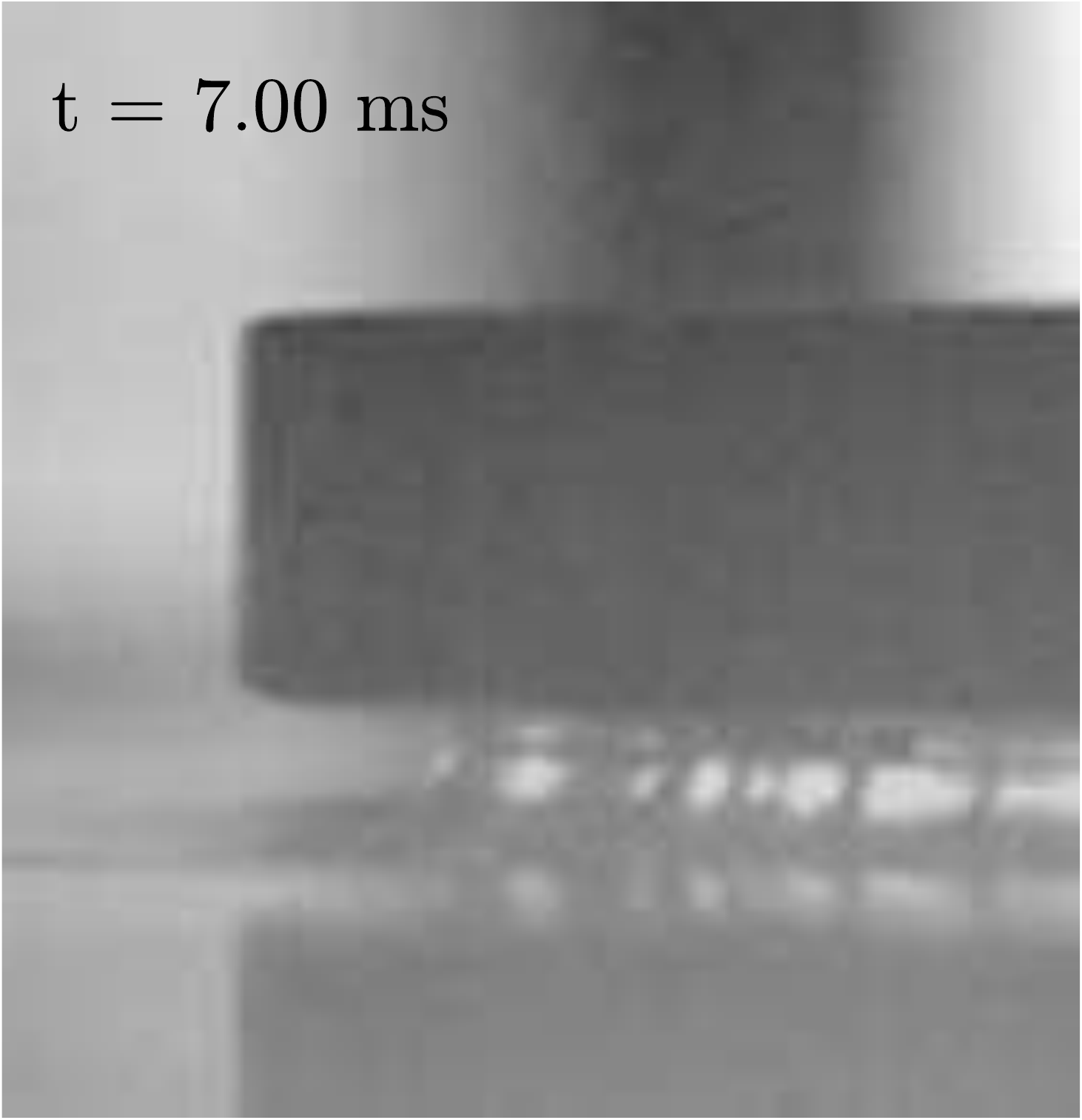}
\caption{\label{fig:4}}
\end{subfigure}

\begin{subfigure}[b]{0.23\textwidth}
\includegraphics[width =\textwidth]{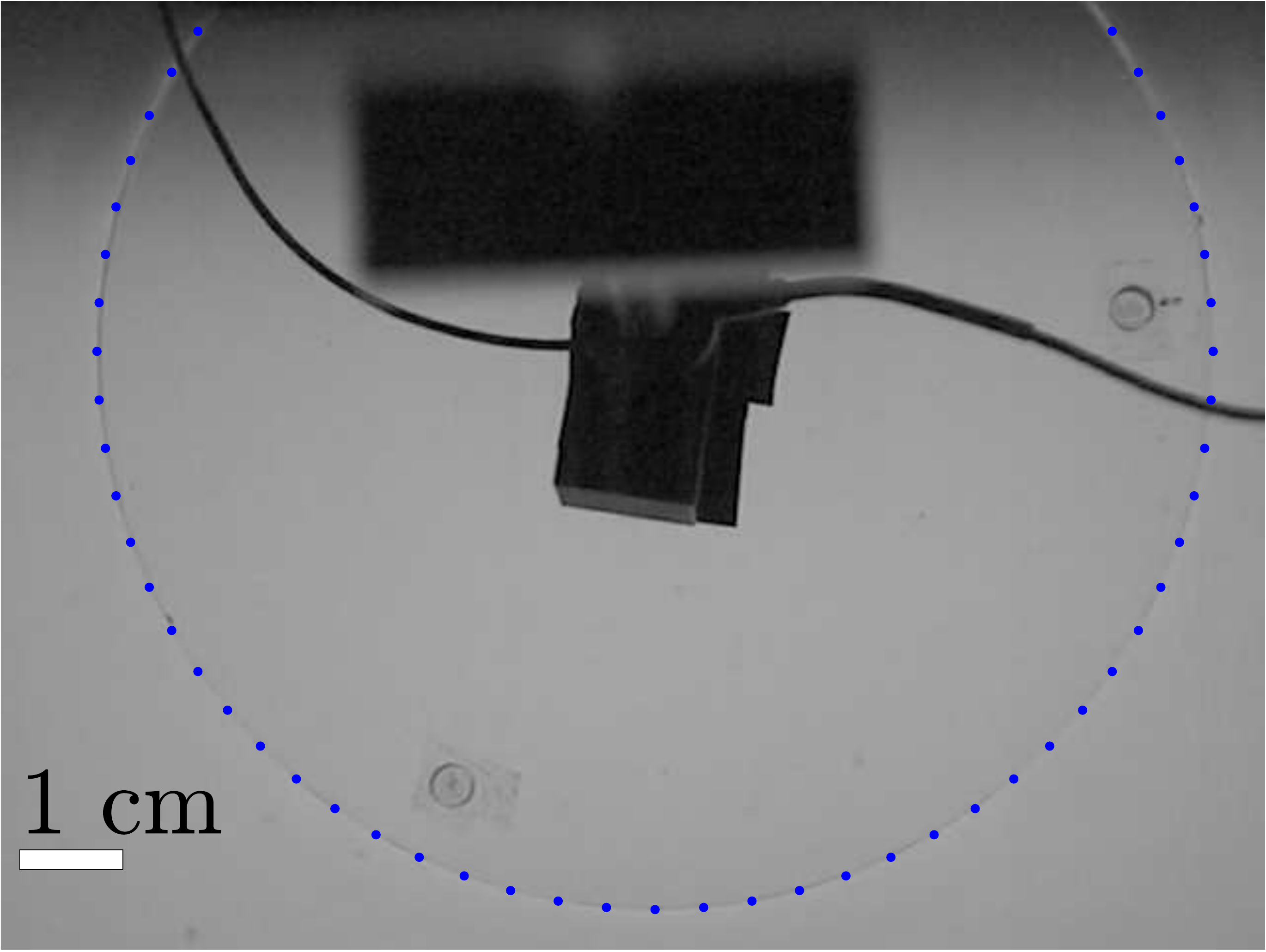}
\caption{\label{fig:5}}
\end{subfigure}
\begin{subfigure}[b]{0.23\textwidth}
\includegraphics[width =\textwidth]{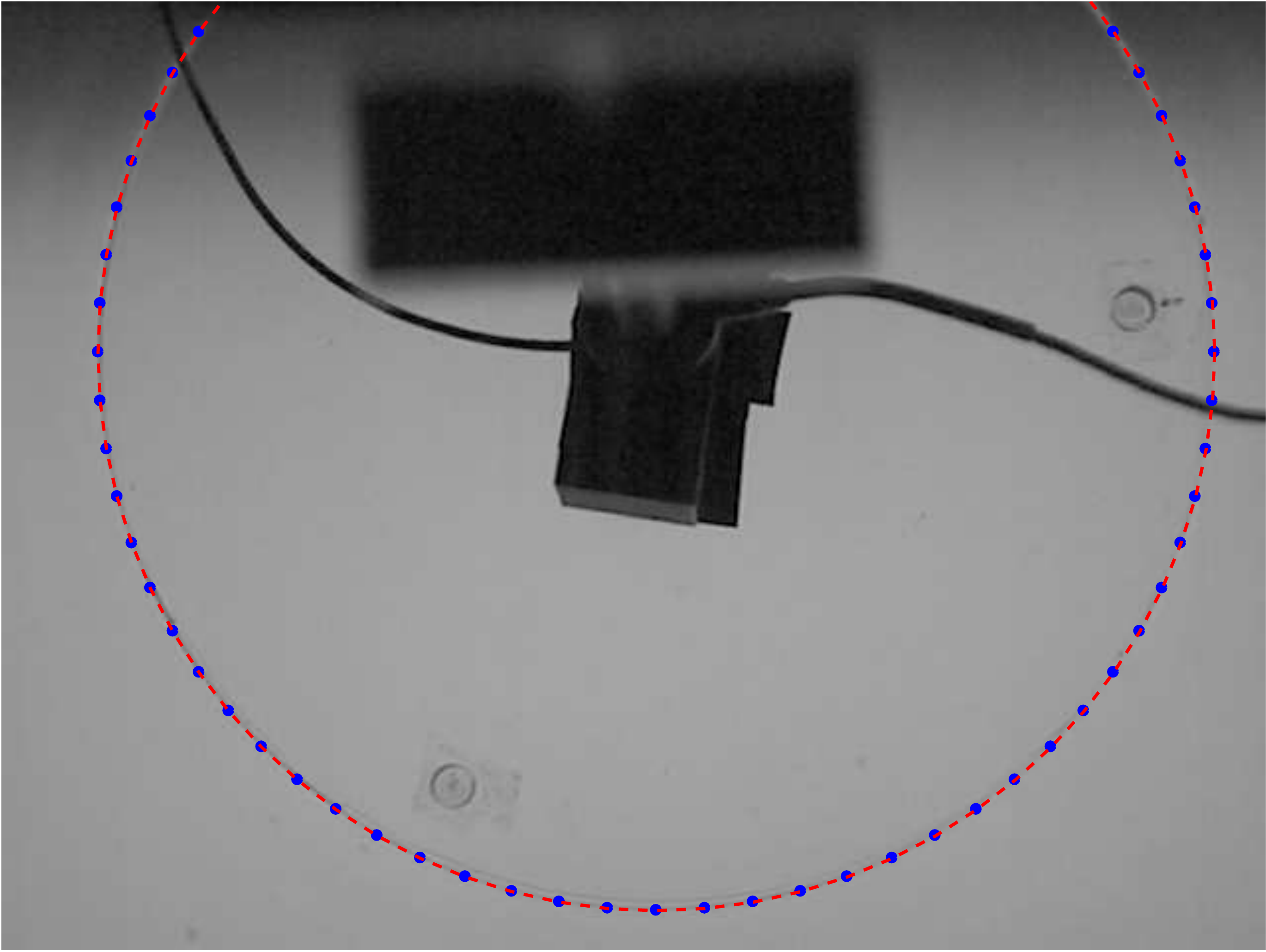}
\caption{\label{fig:6}}
\end{subfigure}
\begin{subfigure}[b]{0.23\textwidth}
\includegraphics[width =\textwidth]{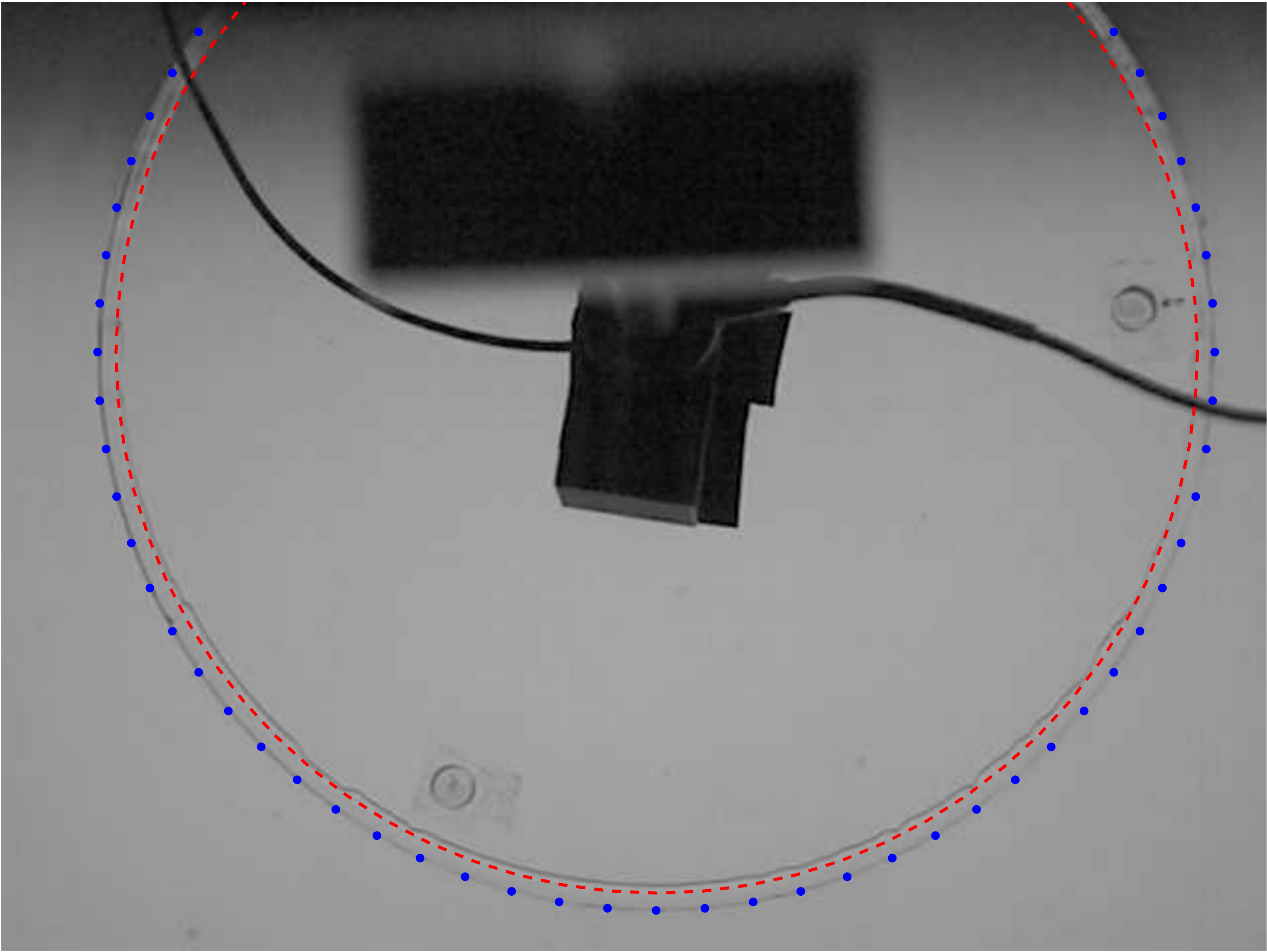}
\caption{\label{fig:7}}
\end{subfigure}
\begin{subfigure}[b]{0.23\textwidth}
\includegraphics[width=\textwidth]{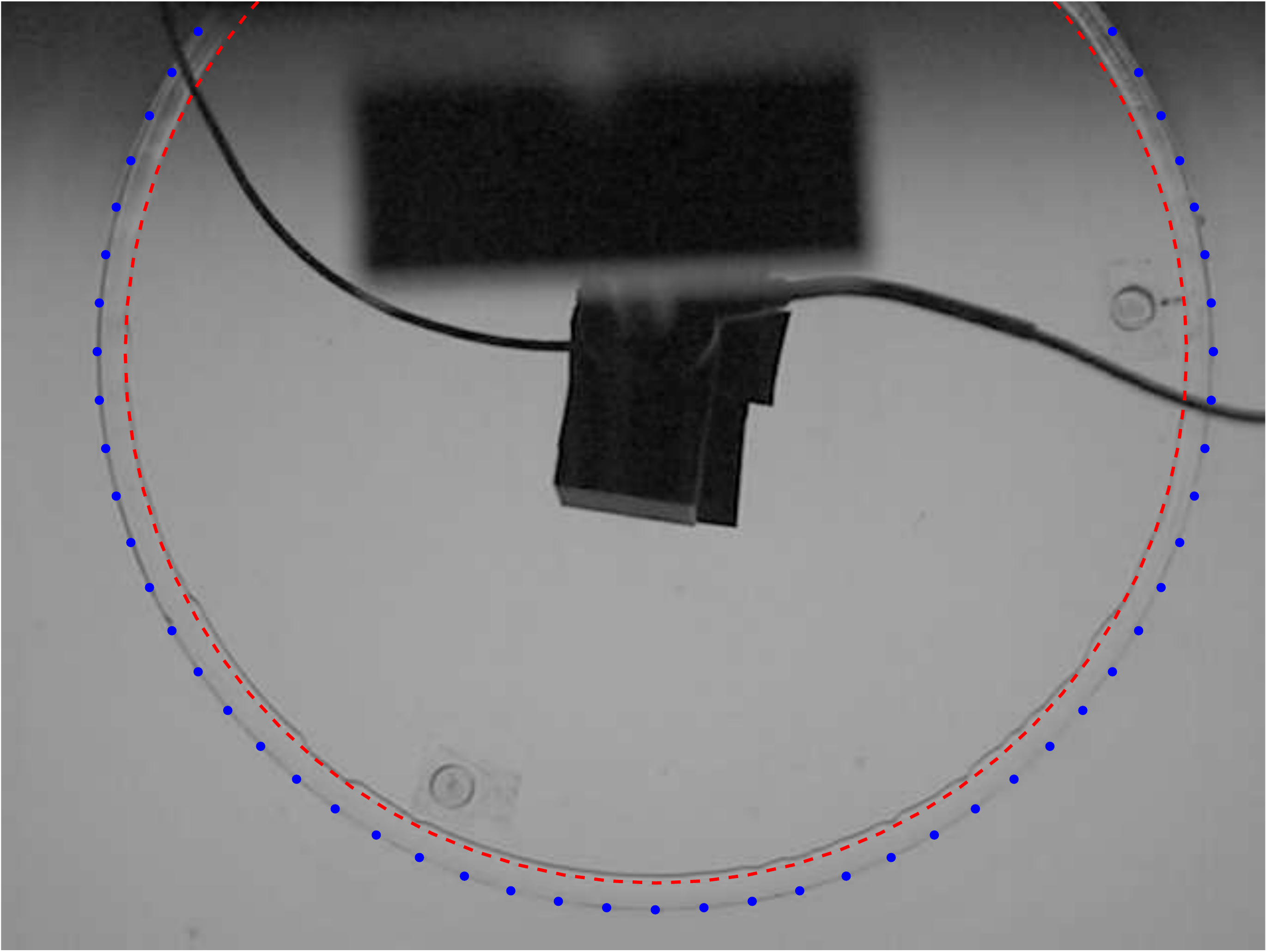}
\caption{\label{fig:8}}
\end{subfigure}
\begin{center}
\includegraphics[width=0.90\textwidth]{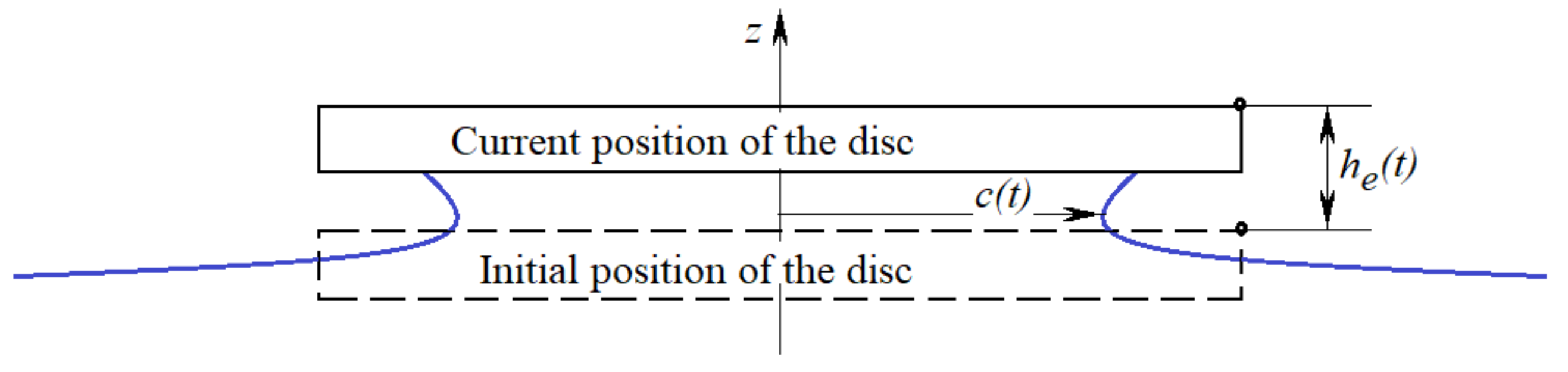}\\
(i)
\end{center}
\caption{\label{fig:high_speed_movies} (a--d) Side view of the edge of the plate and (e--h) top view of the plate where the blue dots mark the diameter of the plate and the red dash traces the detection of the contact line. Images acquired at 15000 fps. (i) Definition of $c(t)$ and $h_e(t)$.}
\end{figure}

The motion of the contact line, in other words the evolution of the wetted area, is recorded by a high-speed camera (NAC Memrecam HX-3) working at 15000 fps. To acquire clear top-view images, we use a LED light (Metaphase Technologies, 9" x 16" White, LED Backlight) at the bottom of the tank, so the perimeter of the disc and the contact line are visible as black lines during the experiment. This lightning configuration is similar to the configuration denoted as ``central LED lighting'' in \citet{Tassin2017}. Moreover, high-speed movies are taken from the side, to compare the evolution of the wetted radius with the distance the plate has lifted. Examples of the images used to measure these quantities are shown in figure \ref{fig:high_speed_movies}, along with a sketch of their definitions.

A few comments should be made here about the image processing techniques used to measure the time evolution of the wetted area. We track the contact line using a custom-made image processing software implemented in Matlab based on the so-called Hough transform \citep{Yuen1990HT}. The Hough transform is a well-known technique used in image analysis to find objects with a given shape, in this case circles, by a voting procedure. In a first step we detect the edges of images like those in figure  \ref{fig:high_speed_movies}(e-h). In this way, two families of pixels are found near the region of interest where the contact line is expected to be found: pixels belonging to the edge of the disc and pixels belonging to the contact line. Each edge point detected adds one vote to all the possible circles that it can belong to. Repeating this procedure for all the edge points, we create a histogram in the parametric space of possible circles, whose local maxima correspond to the circles actually found in the image, i.e. containing a large number of points. In the case of the images shown here, this histogram exhibits two peaks, which correspond to the plate edge and the contact line respectively. We should point out that, using this procedure, it is possible to detect the contact line when it is still very close to the plate edge. Thus, this technique proves to be essential to capture the first instants of the contact line motion.

\section{Comparison of the linearised theory of water exit with experiments}
\label{sec:comparison_linearized_theory}

In this section we present the results of the experiments summarised in table \ref{tab:sessions}. For each realization the accelerometer and load cell provide the acceleration of the center of the plate and the total force pulling it upwards, respectively.
Moreover, the contact line dynamics and the height of the plate during the first instants are obtained through digital image processing of high-speed movies.

The accelerometer measures the acceleration of the plate, $a(t)$, whereas the load cell measures the total force, $F_{exp}(t)$, with which the driving mechanism pulls the instrumented plate of mass $M$. These quantities, together with the hydrodynamic force $F_h(t)$, are related through Newton's second law:
\begin{equation}
    F_{exp} = F_h + Ma.
    \label{eq:def_Fexp}
\end{equation}
From the linearised theory of \cite{Korobkin2013}, $F_h$ is approximated by the product of the disc added mass times its acceleration, thus $F_{exp}$ is expected to be proportional to $a$. More specifically,
\begin{equation}
F_h = m_a \, a, \;\;\;\mathrm{with}\;\;\; m_a = \frac{4}{3}\rho c^3,
\label{eq:Fh_rigid}
\end{equation}
so
\begin{equation}
    F_{exp} = \left(M + \frac{4}{3}\rho c^3\right)a,
    \label{eq:Fexp}
\end{equation}
with $c$ denoting the radius of the wetted area. The added mass term arises from the suction pressure forces that the plate communicates to the parcel of surrounding liquid that follows its accelerated motion. The effect of these forces is equivalent to endow the plate with an additional mass that must be also accelerated with it, thus effectively increasing its apparent inertia. Notice that, in our experiments, this added mass is indeed much larger than the mass of the instrumented disc itself.

\begin{figure}
\centering
\begin{tabular}{lr}
\hspace{-3mm}\includegraphics[width=0.5\textwidth]{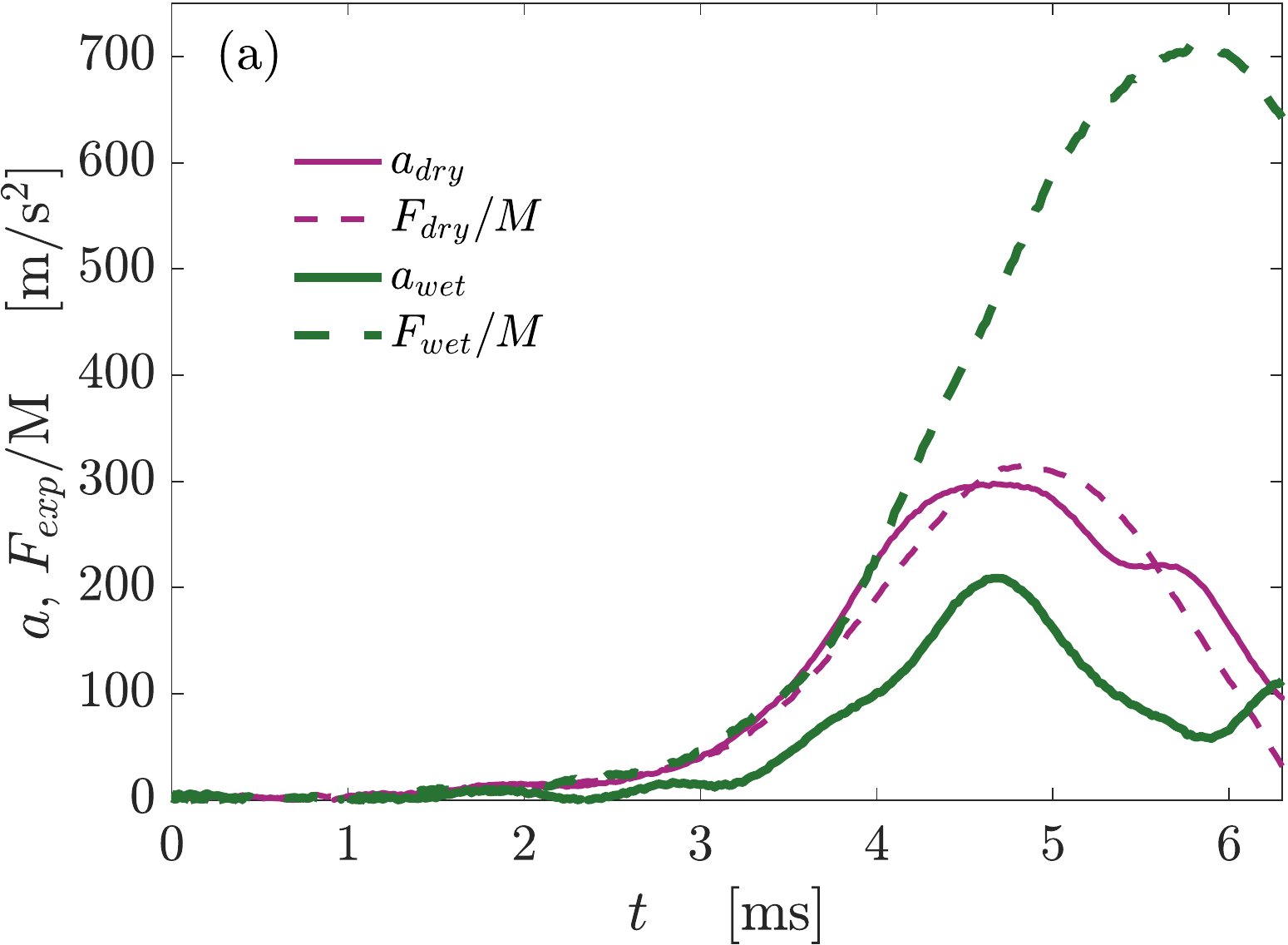} &
\includegraphics[width=0.5\textwidth]{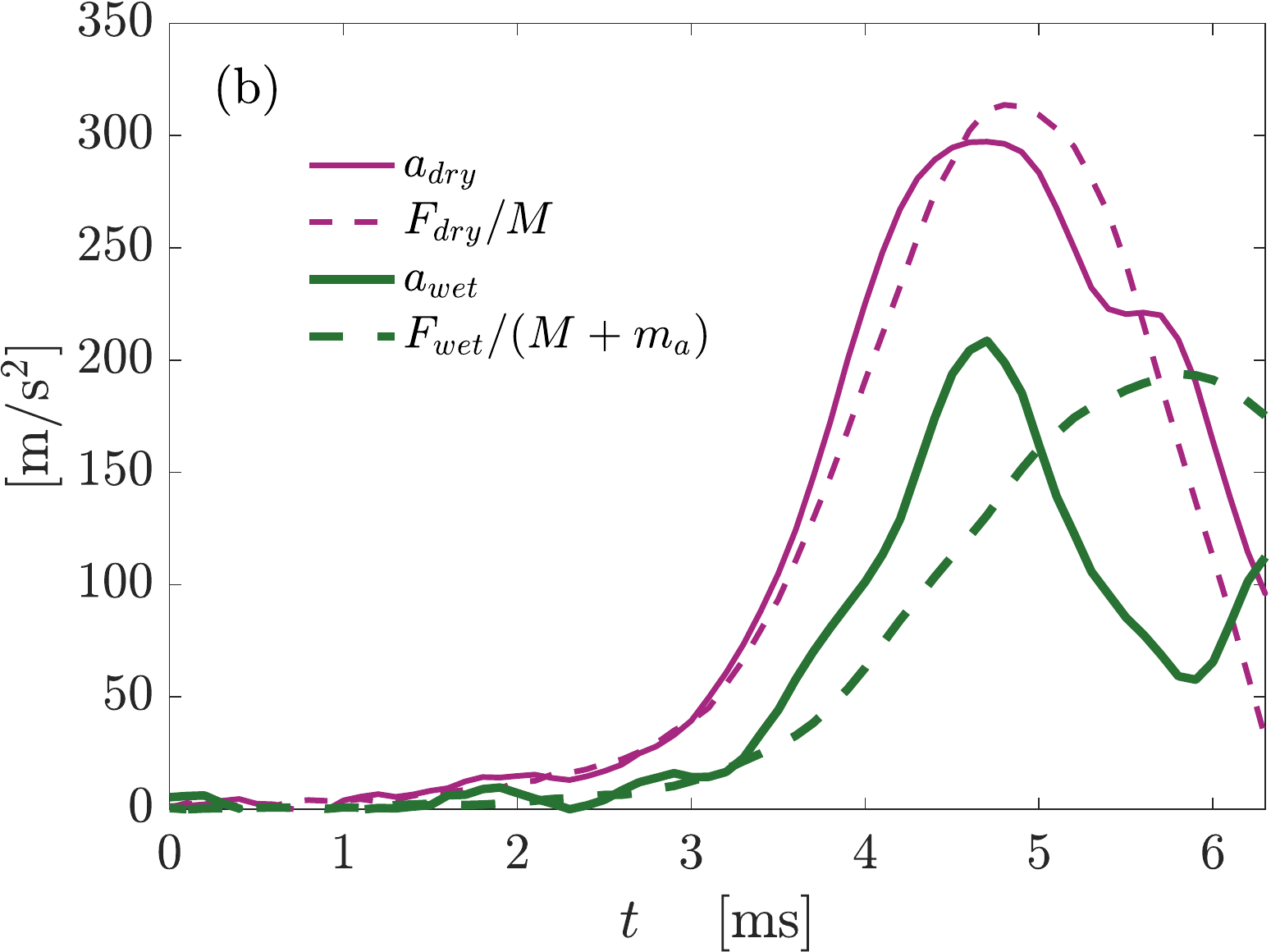}
\end{tabular}
\caption{\label{fig:dry_wet}Comparison between the force and acceleration measurements obtained in a dry and a wet case respectively. Sessions 1 and 2 from table \ref{tab:sessions}.}
\end{figure}

To highlight the relevance of this hydrodynamic suction force, $F_h$, we show in figure \ref{fig:dry_wet}a the force and acceleration measured for the same experimental conditions, but in two realizations with the plate being dry and wet respectively. While in the dry case, $F_h = 0$, the force divided by the mass of the plate coincides with the acceleration within the experimental variability, the acceleration in the wet case is substantially smaller, as a consequence of the hydrodynamic suction force that pulls the plate downwards. It should be remarked that, during the first 4 ms, the force acting on the plate is identical in both the dry and wet cases. Figure \ref{fig:dry_wet}b shows that, taking into account the added mass of the wetted plate, yields a reasonable agreement with equations (\ref{eq:def_Fexp}) and (\ref{eq:Fexp}) during the initial period of 4 ms.

\begin{figure}
\centering
\begin{tabular}{lr}
\hspace{-3mm}\includegraphics[width=0.5\textwidth]{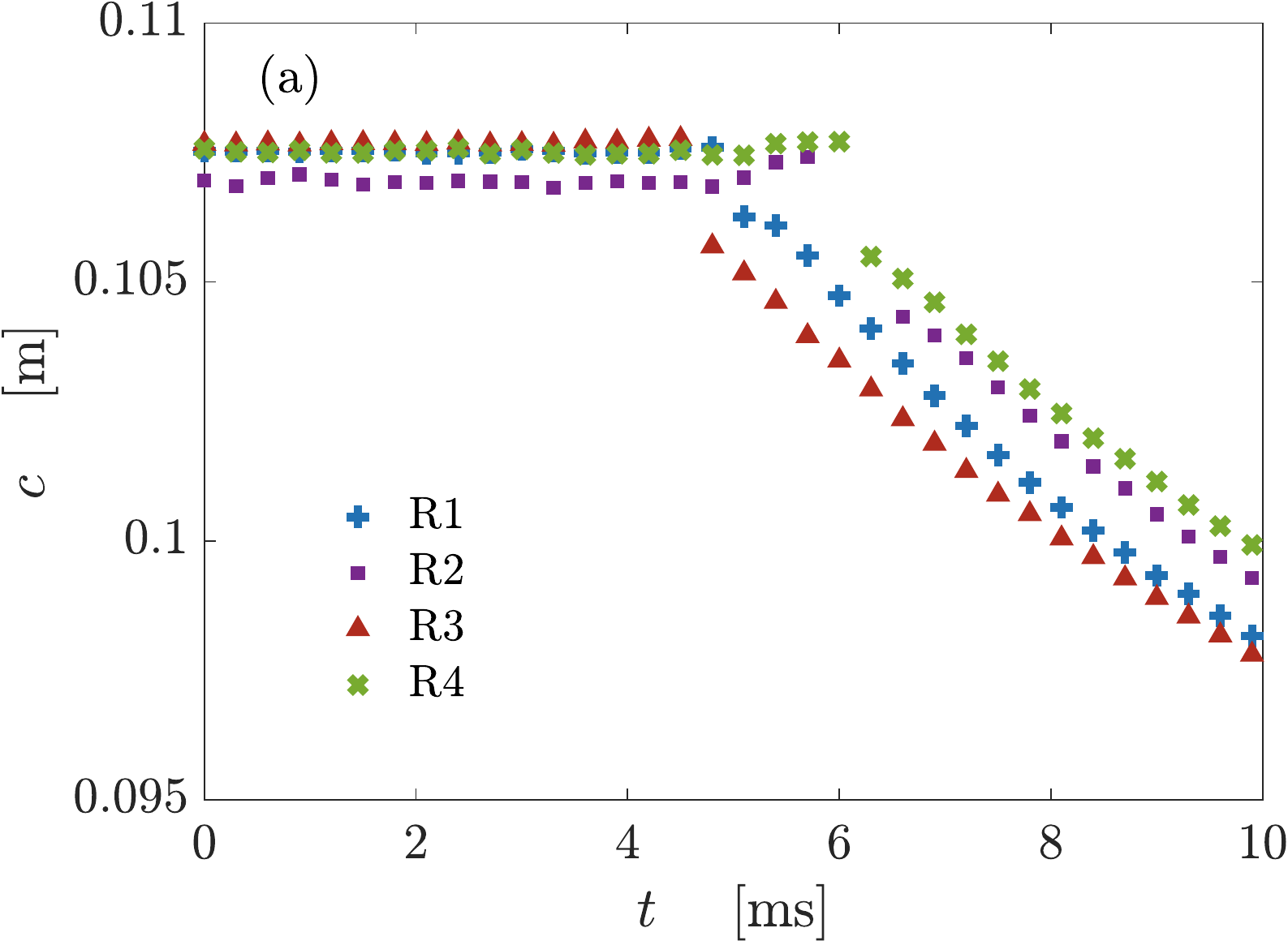} &
\includegraphics[width=0.48\textwidth]{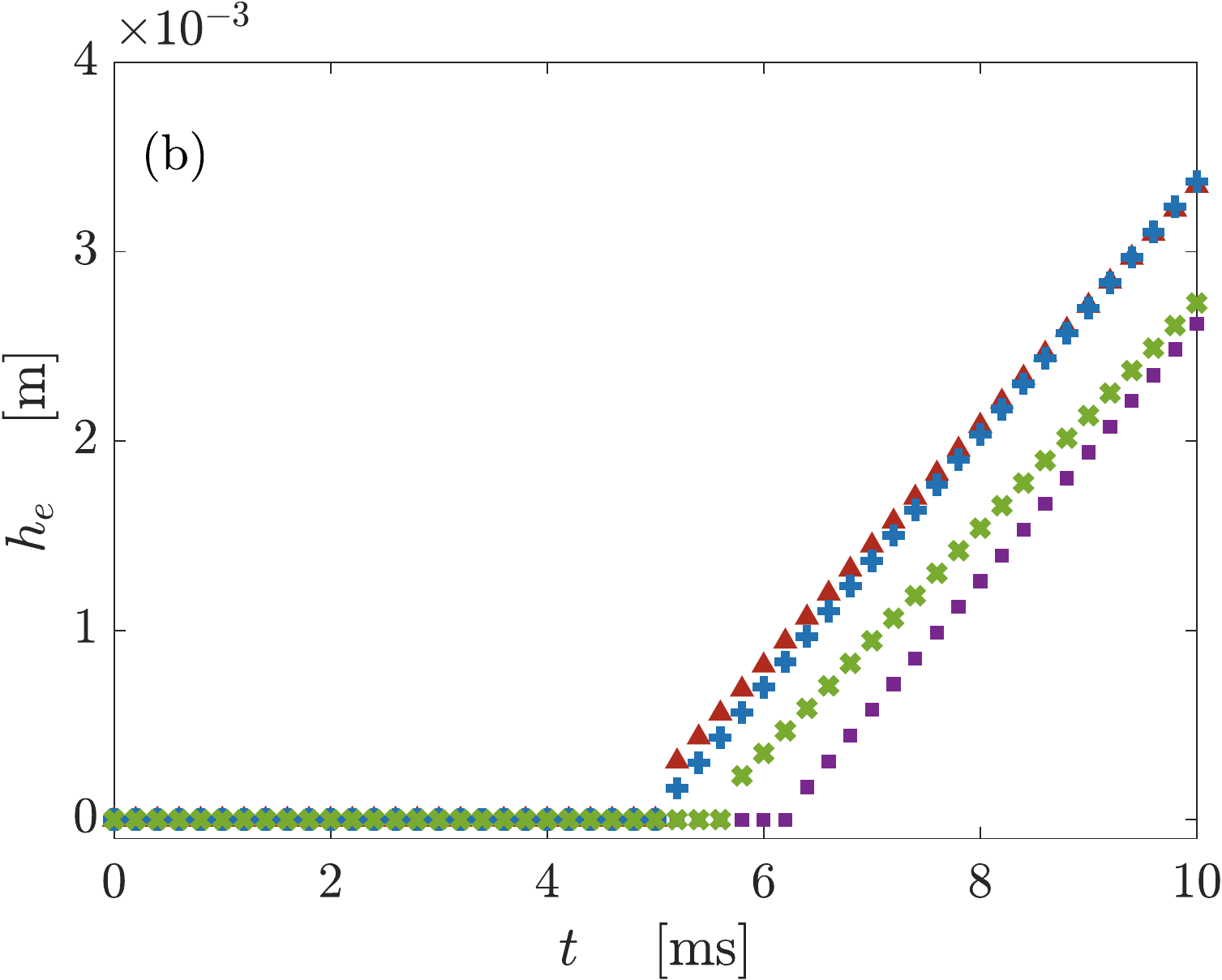}
\end{tabular}
\caption{\label{fig:wetted_radius} Evolution of the radius of the wetted area and the vertical displacement of the edge of the plate for four experimental realizations of session 3.}
\end{figure}

It should be pointed out that the radius of the wetted area, $c(t)$, as well as the forces and the plate's acceleration, are functions of time. However, in practice it is possible to neglect the recoil of the wetted area, since its radius, $c$, barely departures from that of the disc, $R$, along the duration of the experiment, always shorter than 10 ms. This is supported by the measurements of the evolution of the wetted radius shown in figure \ref{fig:wetted_radius}. Figure \ref{fig:wetted_radius}a does not show any motion of the wetted area during the first 4 ms. Indeed, until the plate moves of the order of the capillary length (around 2 mm, see figure \ref{fig:wetted_radius}b) the contact line recoils a very little distance. In order to further prove this, we record movies of the side view of the experiment focusing on the edge (Figure \ref{fig:high_speed_movies}a-d). Notice that only after 4 ms the motion of the contact line can be determined from the top-view, when the contact line detaches from the edge and slides along the lower surface of the plate. In summary, these observations support that $c(t)$ can be approximated as $R$ in the calculations of the hydrodynamic force.

\begin{figure}
\centering
\includegraphics[width=0.55\textwidth]{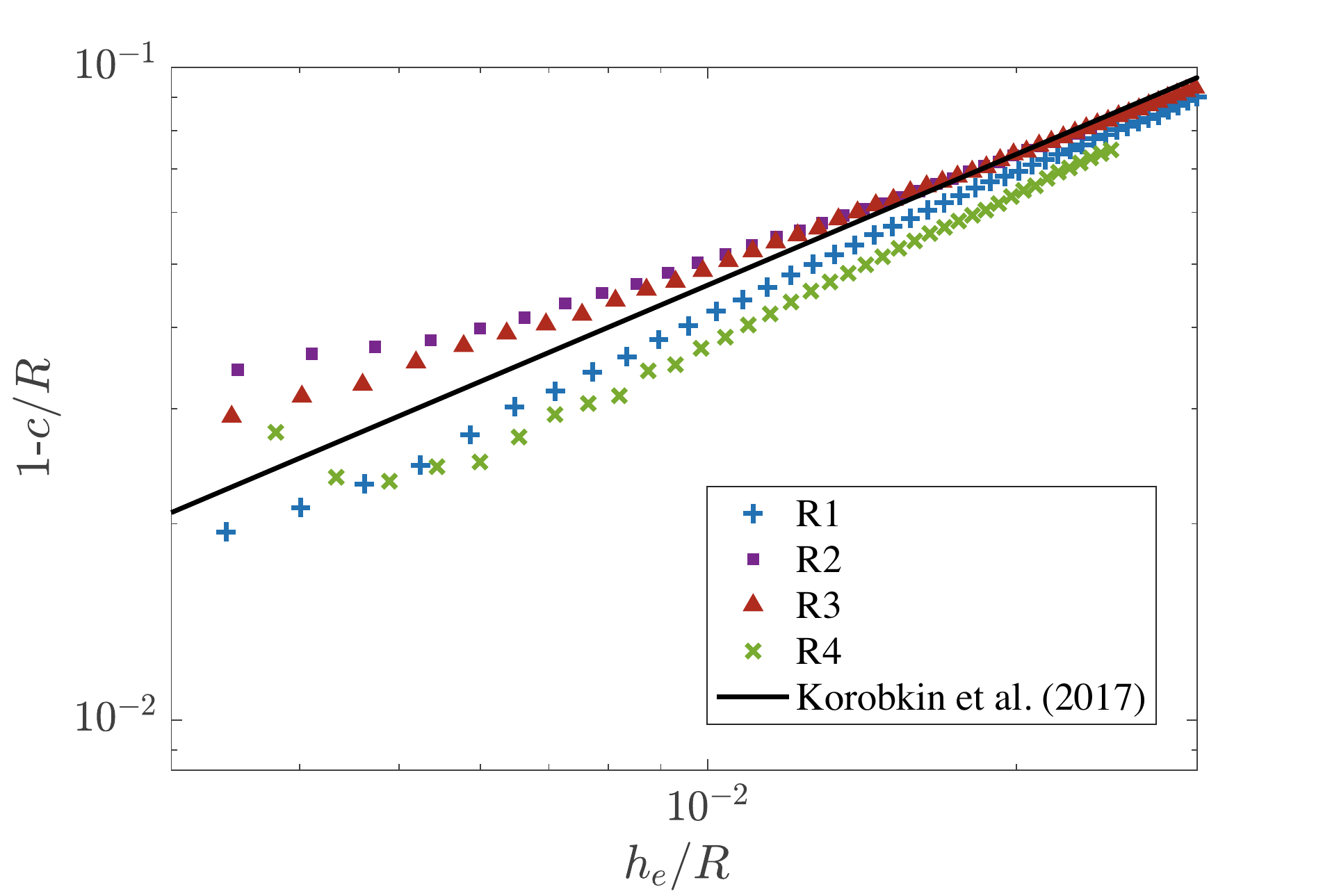}
\caption{\label{fig:self_similar_solution}Log-log plot of $1-c/R$ versus $h_e/R$, showing the convergence to the law $1-c/R \sim (h_e/R)^{2/3}$ (solid line) predicted by \cite{KorKhaba2017}.}
\end{figure}

At longer times, when the contact line detaches from the edge and recoils, its position can be related to the instantaneous displacement of the plate edge, $h_e(t)$, using the ideas of \cite{KorKhaba2017}. In that paper, the authors found a self-similar solution to describe the dynamics of the free surface close to the edge of a plate lifted from the water surface at a large constant acceleration. This solution predicts that the contact line displaces with time a distance $\Delta c \sim t^{4/3}$. On the other hand, since the acceleration was assumed constant, then $h_e \sim t^2$. Combining these estimates, we get $\Delta c \sim h_e^{2/3}$. Even though the acceleration is not constant in our experiments, this power law is indeed recovered once the contact line starts to slide along the plate, as illustrated in figure \ref{fig:self_similar_solution}.

Figure \ref{fig:dry_wet}b sheds light on the dynamics of the coupled water-disc system. Although at short times the acceleration compares reasonably well with the prediction derived from equations (\ref{eq:def_Fexp}) and (\ref{eq:Fh_rigid}), at about $t \approx 4.5$ ms the behavior of the acceleration changes dramatically. In particular it decays, which is in clear contradiction with these equations, which predict a linear relation between $F_{exp}$ and $a$. They are not proportional any more, but one decays whereas the other grows monotonically.

In the next sections we will show that the prediction of proportionality, equation (\ref{eq:Fexp}), is based on the assumption that the plate is rigid. Once elastic effects are taken into account, the oscillatory behavior of the acceleration shown in figure \ref{fig:dry_wet}b --and confirmed by dedicated experiments-- can be explained and quantitatively predicted.

\section{Elastic effects on water exit}\label{elastic}

In order to explain the experimental results of the previous section, we assume that the hydroelastic interaction of the acrylic disc and the liquid is important during the early stage. This assumption is not obvious because the disc does not demonstrate visible elastic motions in the experiments. To prove that the disc elasticity is important in the water exit process, we generalize the linearised model of water exit of \cite{Korobkin2013} by including the elasticity of the lifted body, and compare the obtained theoretical results with the experimental ones. The theoretical analysis is limited to axisymmetric problems, where the contact line remains attached to the edge of the plate, which corresponds to the conditions of the water exit in present the experiments during the early stage.

\subsection{Formulation of the axisymmetric exit problem and its solution}\label{3-1}

The problem of water exit is formulated in cylindrical coordinates $(r, z)$, where $r$ is the radial coordinate and $z$ is the vertical one, which coincides with the symmetry axis of the plate. The plane $z=0$ corresponds to the initial position of both the lower surface of the disc and the liquid surface, with the liquid occupying the lower half-space, $z<0$. The disc touches the water surface with zero draft. Thus the initial position of the lower surface of the disc is $z=0$, $r<R$. Then the disc is moved suddenly upwards by an external force applied at the disc centre. The position of the lower surface of the moving disc is described by the equation $z=w(r,t)$, where $w(r,t)$ accounts for both the rigid and elastic components of the disc displacement. During the early stage, when displacements are small compared to the disc radius $R$, the flow equations and the boundary conditions can be linearised. In particular, this means that the boundary conditions can be imposed on the initial position of the liquid boundary.

Under the assumptions of the early stage, the flow caused by lifting an elastic disc from the water surface is described by a velocity potential, $\varphi(r,z,t)$, which satisfies Laplace equation,
\begin{equation}
\frac{1}{r}\frac{\d}{\d r}\left( r \frac{\d\varphi}{\d r}\right) +\frac{\d^2\varphi}{\d z^2}=0 \qquad (z<0),
\label{eq:laplace_equation}
\end{equation}
in the initial flow region, the linearised dynamic condition on the free surface,
\begin{equation}
\vp=0 \qquad (z=0, \quad r>R),
\end{equation}
and the body boundary condition on the disc,
\begin{equation}
\frac{\d\vp}{\d z}=\frac{\d w}{\d t} \qquad (z=0,\quad  r<R).
\label{eq:boundary_condition_disc}
\end{equation}
Furthermore, this potential decays at infinity, as $r^2+z^2\to\infty$.

The disc displacement, $w(r,t)$,  is governed by the Bernoulli-Euler equation of thin elastic plates,
\begin{equation}
m\frac{\d^2 w}{\d t^2} +D\nabla^4 w= p_{ext}(r,t) +p_{h}(r,0,t) \qquad (r<R, \quad t>0),
\label{eq:plate_equation}
\end{equation}
where $m$ is the mass of the disc per unit area, $m=\rho_p h_p$, $D$ the rigidity coefficient, $D=Eh_d^3/[12(1-\nu^2)]$, for an elastic disc of constant thickness, see Table \ref{tab:properties} for the elastic constants of the disc material. The hydrodynamic pressure, $p_h(r, 0, t)$, acting on the disc/water interface, $z=0$, is given by the linearised Bernoulli equation,
\begin{equation}
    p(r, 0, t) =-\rho\vp_t (r, 0, t),    
\end{equation}
and $p_{ext}(r,t)$ is the external load caused by the driving mechanism, see figure \ref{fig:experimental_setup}, and acting on the disc through the load cell and the connector. The displacement $w(r, t)$ is positive where the disc moves upwards. Note that different parts of the disc can move in different directions. For example, the centre of the disc may move upwards but the edge of the disc moves downwards at the same time.

The connector is not axisymmetric in experiments. Therefore, the external load and the disc deflection should depend on the azimuthal coordinate, strictly speaking. We neglect this effect in the present analysis assuming that the external load is uniformly distributed over a small circular area of radius $r_c$, $p_{ext}(r,t)=P(t)$, where $r<r_c$, with $F_{ext}(t)=\pi r_c^2 P(t)$ being the total force acting on the disc.

In the experiments, the weight of the connector, $M_c$, which is placed  in between the load cell and the disc, is comparable with the weight, $M_p=\pi R^2 m$, of the disc. To account for this extra weight, Newton's second law for the connector,
\begin{equation}
M_c \frac{\d^2 w}{\d t^2}(0, t)=F_{exp}(t)-F_{ext}(t),    
\end{equation}
is used, where $F_{exp}(t)$ is the force measured by the load cell, and $-F_{ext}(t)$ is the force acting on the connector from the disc. We assume that the connector is rigidly connected to the centre of the disc, thus the displacement of the connector is approximately $w(0,t)$. Then the external load can be approximated by
\begin{equation}
p_{ext}(r, t)=\left(F_{exp}(t)-M_c \frac{\d^2 w}{\d t^2}(0, t)\right) \frac{\delta(r)}{2\pi r},
\label{eq:definition_pext}
\end{equation}
where $\delta(r)$ is the Dirac delta function. In the present model, the force $F_{exp}(t)$ is assumed known from experimental measurements.

The disc edge, $r=R$, is free of stresses and shear forces. The radial bending moment and the Kelvin-Kirchhoff edge reaction are zero at the edge. For axisymmetric deflections of a circular plate, these two conditions read
\begin{equation}
\frac{\d^2 w}{\d r^2}+\frac{\nu}{r}\frac{\d w}{\d r}=0, \qquad \frac{\d}{\d r}\left(\frac{1}{r}\frac{\d}{\d r}\left( r\frac{\d w}{\d r}\right)\right)=0 \quad (r=R).
\end{equation}

Initially, $t=0$, both the disc and the water are at rest,
\begin{equation}
w(r,0)=0, \quad w_t(r,0)=0, \quad \vp(r, z, 0)=0.
\label{eq:initial_conditions}
\end{equation}
The problem formulated by equations (\ref{eq:laplace_equation})-(\ref{eq:initial_conditions}) is coupled, the hydrodynamic loads and the disc displacement should be determined at the same time. The problem is solved by the normal mode method \citep{KhabakhpashevaKorobkin1998, Korobkin2000, Khabakhpasheva2006, Khabakhpasheva_etal2013}. Within this method the disc displacement is sought in the form
\begin{equation}
w(r,t)=h(t)+\sum_{n=1}^{\infty}a_n(t) W_n(\rt),
\label{eq:series_definition}
\end{equation}
where $\rt=r/R$ is the non-dimensional radial coordinate, $\rt<1$, $a_n(t)$ are the principal coordinates of the elastic deflections of the disc, which are to be determined, and $h(t)$ is the rigid-body displacement of the disc. The functions $W_n(\rt)$ are the non-trivial bounded solutions to the homogeneous boundary value problem
\begin{equation}
\nt^4 W_n=k_n^4 W_n \quad (\rt<1),
\label{eq:eigenvalue_equation}
\end{equation}
\begin{equation}
W_n''+\nu W_n'=0, \quad (\nt^2 W_n)'=0 \quad (\rt=1),    
\end{equation}
where
\begin{equation}
\nt^2 W_n= \frac{1}{\rt}\frac{\d}{\d \rt}\left( \rt\frac{\d W_n}{\d \rt}\right),    
\end{equation}
a prime stands for the derivative in $\rt$, and $k_n$ are the corresponding eigenvalues. The functions $W_n(\rt)$, known as normal modes of the free-free circular elastic disc,  describe the axisymmetric shapes of free vibrations of a circular disc with its edge being free of forces and bending stresses with frequencies proportional to $k_n^2$ \citep{Leissa1969}. The normal modes are orthogonal and read
\begin{equation}
W_n(\rt)=A_n\left( {\rm J}_0(k_n\rt) -\frac{\rJ_1(k_n)}{\rI_1(k_n)}\rI_0(k_n\rt)\right),
\label{eq:normal_modes_full_expression}
\end{equation}
where $k_n$ , $n\geq 1$, are the positive solutions of the equation
\begin{equation}
\frac{\rJ_1(k_n)}{\rJ_0(k_n)}+\frac{\rI_1(k_n)}{\rI_0(k_n)}=\frac{2(1-\nu)}{k_n},
\end{equation}
and the coefficients $A_n$ are determined by the normalization condition,
\begin{equation}
\int_0^1 W_n(\rt) W_m(\rt) \rt \rd \rt=\delta_{nm},
\label{eq:normalization_condition}
\end{equation}
$\delta_{nm}=0$ for $n\neq m$ and $\delta_{nn}=1$. Equations (\ref{eq:normal_modes_full_expression})-(\ref{eq:normalization_condition}) give
\begin{equation}
A_n=\left(\rJ_0^2(k_n)-\frac{2\nu(1-\nu)}{k_n^2}\rJ_1^2(k_n)-\frac{2(1-\nu)}{k_n}\rJ_0(k_n)\rJ_1(k_n)\right)^{-\frac{1}{2}}.    
\end{equation}
Note that
\begin{equation}
\int_0^1 W_n(\rt)\rt \rd \rt=0,
\label{eq:Wn_integral_zero}
\end{equation}
which means that the elastic modes with $n\geq 1$ are orthogonal to the rigid mode, $W_0(\rt)=\sqrt{2}$, which corresponds to $k_0=0$.

The solution of the hydrodynamic problem (\ref{eq:laplace_equation})-(\ref{eq:boundary_condition_disc}) is given by
\begin{equation}
\vp(R\rt, 0,t)=\int_{\rt}^1\frac{\chi(\mu, t)\rd\mu}{\sqrt{\mu^2-\rt^2}}, \quad \chi(\mu, t)=\frac{2R}{\pi}\int_0^{\mu} \frac{w_t(R\sigma,t)\sigma\rd\sigma}{\sqrt{\mu^2-\sigma^2}},
\label{eq:solution_generic}
\end{equation}
see Appendix A in \cite{KorobkinScolanJFM2006}. It is convenient to introduce the functions $Q_n(x)$ and the coefficients $W_{nk}$ by
\begin{equation}
Q_n(x)=\frac{1}{x}\int_0^x \frac{W_n(\sigma)\sigma \rd\sigma}{\sqrt{x^2-\sigma^2}}, \quad W_{nk}=\frac{2}{\pi}\int_0^1x^2Q_n(x)Q_k(x)\rd x,
\label{eq:definition_Qn_Wnk}
\end{equation}
see equations (3.19) and (3.37) in \cite{Pegg_etal2018}. In particular, $Q_0(x)=\sqrt{2}$. The functions $Q_n(x)$ and the coefficients $W_{nk}$ are expressed through the Bessel, trigonometric and hyperbolic functions similar to those in \cite{Pegg_etal2018}, Appendixes B and C, where the corresponding integrals were evaluated for a simply supported circular elastic disc.

Substituting (\ref{eq:series_definition}) in (\ref{eq:solution_generic}) and using (\ref{eq:definition_Qn_Wnk}) we find the velocity potential on the disc surface,
\begin{equation}
\chi(\mu, t)=\frac{2R}{\pi}\mu\left(h'(t)+\sum_{n=1}^{\infty}a_n'(t)Q_n(\mu)\right),    
\end{equation}
\begin{equation}
\vp(R\rt, 0,t)=\frac{2R}{\pi}\left(h'(t)\sqrt{1-\rt^2}+\sum_{n=1}^{\infty}a_n'(t)\int_{\rt}^1 \frac{\mu Q_n(\mu)\rd\mu}{\sqrt{\mu^2-\rt^2}}\right).    
\end{equation}
The latter equation provides the asymptotic behaviour of the radial flow velocity near the edge of the disc,
\begin{equation}
\frac{\d \vp}{\d r}(r, 0,t)\sim -\frac{2R V(t)}{\sqrt{R^2-r^2}}, \quad V(t)=h'(t)+\sum_{n=1}^{\infty}a_n'(t)Q_n(1).
\label{eq:radial_velocities}
\end{equation}

Multiplying both sides of the plate equation (\ref{eq:plate_equation}) by $\rt$ and integrating in $\rt$ from 0 to 1 using (\ref{eq:series_definition}), (\ref{eq:definition_pext}) and (\ref{eq:Wn_integral_zero}), we arrive at the following equation for the rigid displacement of the disc,
\begin{equation}
h''(t)=\frac{F_{exp}(t)}{M+m_a}-\sum_{n=1}^{\infty}a_n''(t)\left\{\frac{M_c}{M+m_a}W_n(0)+\frac{3\pi}{2\sqrt{2}}\frac{m_a}{M+m_a}W_{n0}\right\},
\label{eq:acceleration_h2}
\end{equation}
where $M=M_c+M_p$ is the mass of the equipped disc, and $m_a$ is the added mass of the disc (equation (\ref{eq:Fh_rigid})). Equation (\ref{eq:acceleration_h2}) provides the acceleration at the disc centre, $a(t)$, which is measured in the experiments,
\begin{equation}
\begin{split}
a(t)= \frac{\d^2 w}{\d t^2}(0, t)=h''(t)+\sum_{n=1}^{\infty}a_n''(t)W_n(0)=\frac{F_{exp}(t)}{M+m_a} + \\
\sum_{n=1}^{\infty}a_n''(t)\left\{\frac{M_p+m_a}{M+m_a}W_n(0)-\frac{3\pi}{2\sqrt{2}}\frac{m_a}{M+m_a}W_{n0}\right\}.
\end{split}
\label{eq:theoretical_center_acceleration}
\end{equation}

Therefore, the elastic deflection of the disc may significantly affect the acceleration of its center.
Formula (\ref{eq:theoretical_center_acceleration}) is reduced to (\ref{eq:Fexp}) when the elastic accelerations are negligible, i.e. $a_k(t) \approx 0$.

The equations for elastic deflections of the disc follow from the plate equation (\ref{eq:plate_equation}). Multiplying both sides of the plate equation (\ref{eq:plate_equation}) by $\rt W_k(\rt)$ and integrating in $\rt$ from 0 to 1 using (\ref{eq:series_definition}), (\ref{eq:definition_pext}), (\ref{eq:eigenvalue_equation}) and (\ref{eq:normalization_condition}), we arrive at the following equations for the principal coordinates $a_k(t)$, $k \ge 1$,
\begin{equation}
a_k''(t)+\omega_k^2 a_k =F_{exp}(t) f_k +\sum_{n=1}^{\infty}a_n''(t) S_{kn},
\label{eq:system_ak}
\end{equation}
where
\begin{equation}
\omega_n^2=\frac{Dk_n^4}{m R^4}, \quad \gamma=\frac{3\pi}{2\sqrt{2}}\frac{m_a}{M+m_a}, \quad
f_k=\frac{M_p+m_a}{2M_p (M+m_a)}\left(W_k(0)-\gamma W_{k0}\right),    
\end{equation}
\begin{equation}
\begin{split}
S_{kn}=S_{nk}=\frac{3\pi}{4\sqrt{2}}\frac{m_a}{M_p}\gamma W_{k0}W_{n0} - \frac{M_c}{2M_p}\frac{M_p+m_a}{M+m_a}W_k(0) W_n(0) \\
+\frac{\gamma M_c}{2M_p}\left( W_k(0) W_{n0}+W_n(0) W_{k0}\right)-\frac{3\pi}{4}\frac{m_a}{M_p}W_{nk}.
\end{split}
\end{equation}
The system (\ref{eq:system_ak}) is integrated in time numerically subject to the initial conditions
\begin{equation}
a_k(0)=0, \quad a'_k(0)=0\quad (n\geq 1)
\label{eq:initial_condition_ak}
\end{equation}
and for the forcing function $F_{exp}(t)$ measured in the experiments. Then the acceleration at the center of the disc is calculated using equation (\ref{eq:theoretical_center_acceleration}) and compared with the results gathered by the accelerometer.

\subsection{Comparison between theoretical and experimental results}\label{sec:comparison_experiments_theory}

The computations are performed for the one-mode approximation with $a_k(t)\equiv 0$ for $n\geq 2$. This is because the conditions of the experiments are axisymmetric only approximately. Therefore, we do not expect that including more axisymmetric terms in the series (\ref{eq:series_definition}) would improve the theoretical predictions of the disc acceleration compared to the experimental ones.

\begin{figure}
\centering
\begin{tabular}{rl}
\hspace{-3mm}
\includegraphics[width=0.5\textwidth]{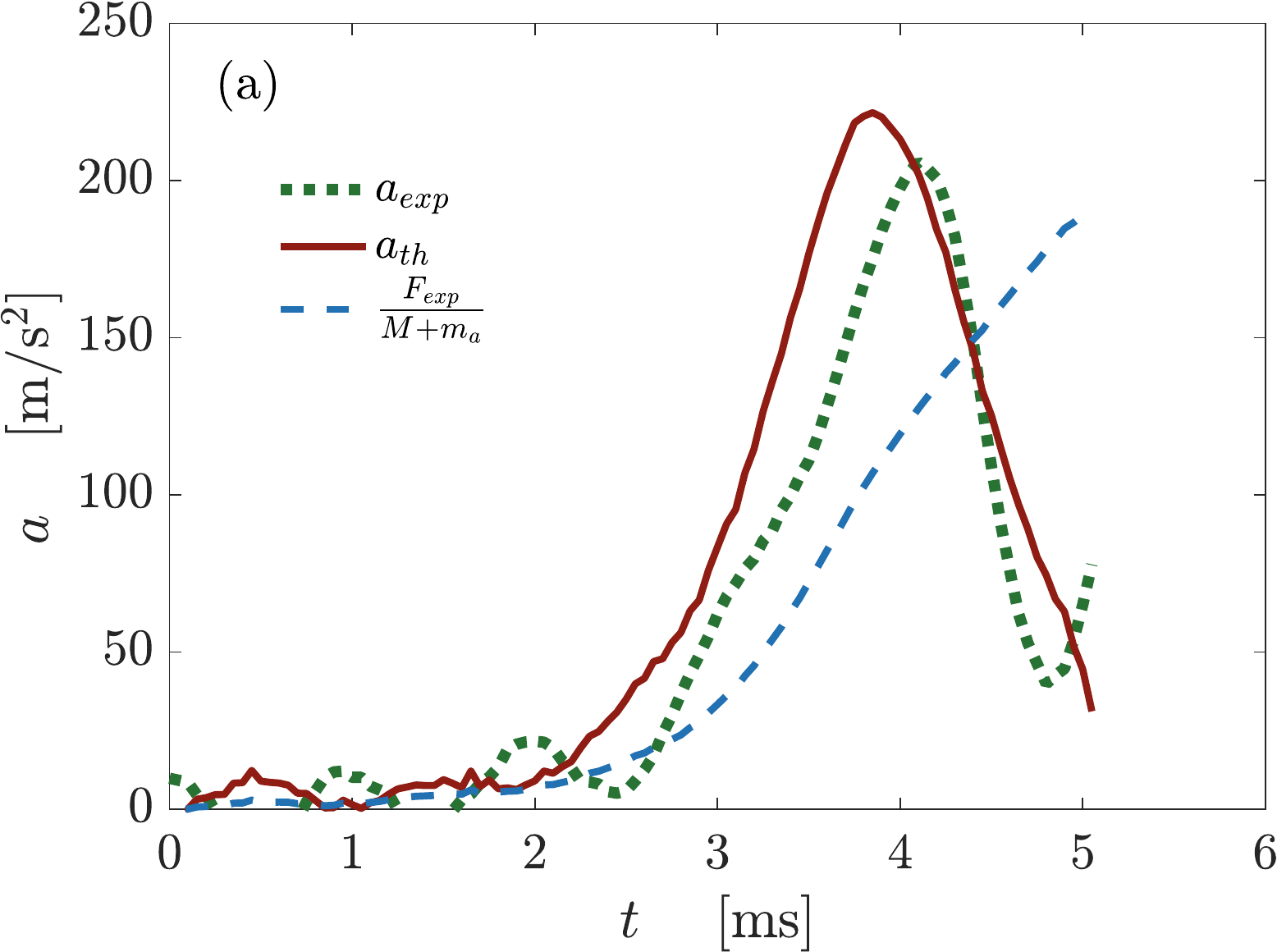} &
\hspace{-2mm}
\includegraphics[width=0.5\textwidth]{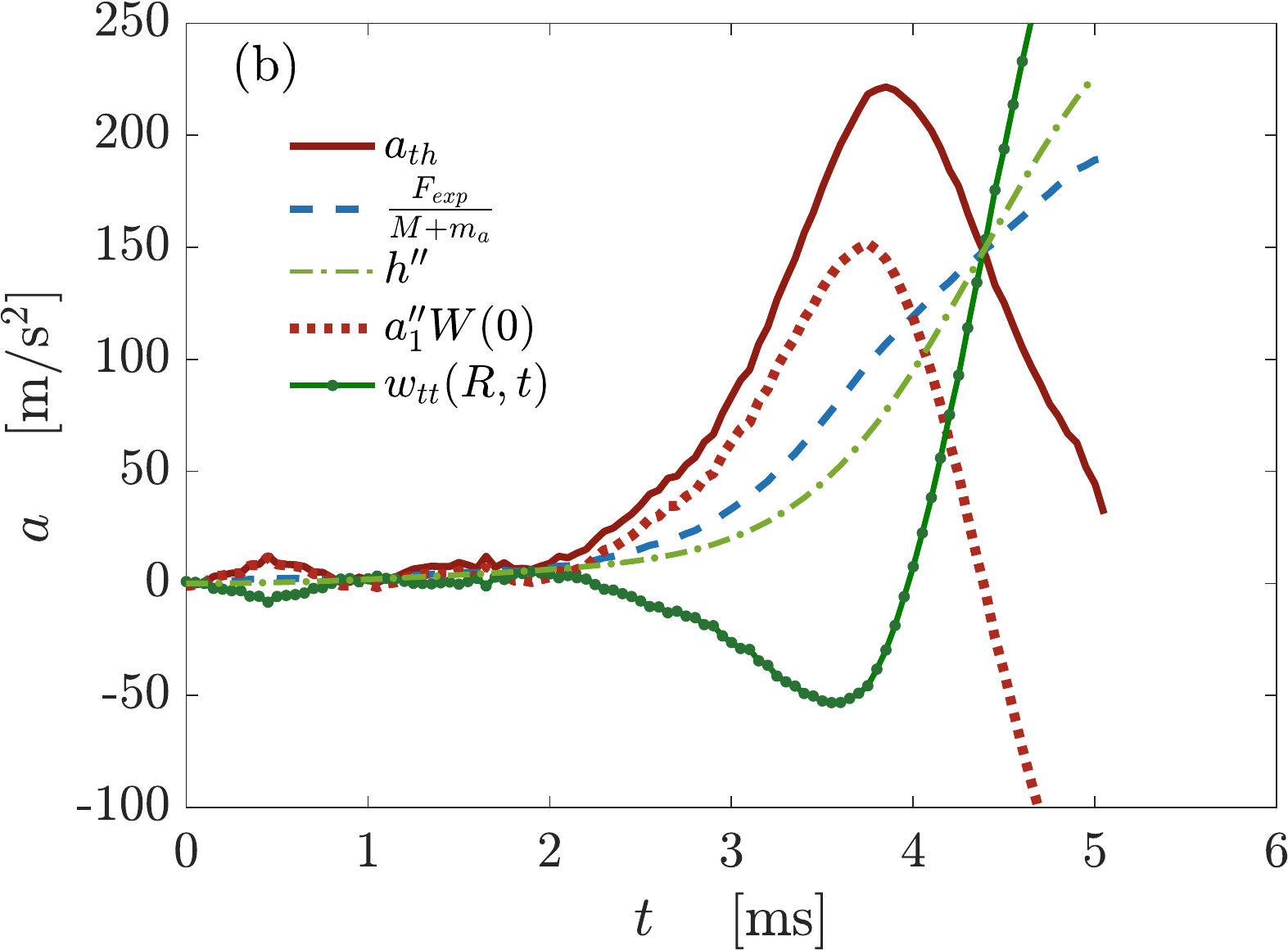}
\end{tabular}
\caption{\label{fig:comparison_acceleration_components}Comparison of experimental and theoretical results for session 3, repetition 1 (see figure \ref{fig:repeatability})}. (a) Dashed line is for normalised excitation force $F_{exp}(t)/(M+m_a)$, solid line is for theoretical acceleration at the disc centre provided by equation (\ref{eq:acceleration_theoretical}), and the dotted line is for the measured acceleration. (b) Normalised excitation force and theoretical acceleration measured at the center of the disc are the same as in (a), dash-dotted line is for the rigid-body acceleration $h''(t)$ given by (\ref{eq:one_mode_h2}), dotted line is for the elastic acceleration at the plate centre, $a''_1W_1(0)$, and the solid line with dots markers are for the acceleration of the disc edge, $w_{tt}(R,t)$.
\end{figure}

For the conditions of the experiments with the plate thickness $h_p = 1$ cm (see table \ref{tab:properties}) we calculate $k_1=3.011$. The frequency and the period of the first dry mode are $\omega_1=3853$ s$^{-1}$ and $T_1=1.63$~ms. (Compare with the corresponding values for the second dry elastic mode, $k_2=6.2$, $\omega_2=16363$ s$^{-1}$, $T_2=0.38$~ms), $\gamma=2.423$, and $W_{10}= 0.143$, $W_1(0)= 2.848$, $W_{11}= 0.218$, $f_1=2.645$ kg$^{-1}$, $S_{11}= -2.921$. The resulting equation for the principal coordinate of the first elastic mode reads
\begin{equation}
a_1''(t)+\Omega_1^2 a_1 = \alpha_1 F_{exp}(t),
\label{eq:a12}
\end{equation}
where the frequency of the first wet mode, $\Omega_1$, and the forcing factor $\alpha_1$ are given by
\begin{equation}
\Omega_1=\frac{\omega_1}{\sqrt{1-S_{11}}}, \qquad \alpha_1=\frac{f_1}{1-S_{11}}.     
\end{equation}
In the conditions of these experiments, $\Omega_1=1946$ sec$^{-1}$ and $\alpha_1=0.675$ kg$^{-1}$. The corresponding period of the first wet elastic mode is equal to $2\pi/\Omega_1=3.223$ ms, which is twice greater than the period of the dry mode, and is comparable with the peak time of the measured acceleration, see figures \ref{fig:comparison_acceleration_components} and \ref{fig:comparison_accelerations}. Equation (\ref{eq:a12}) is integrated numerically for the function $F_{exp}(t)$ measured by the load cell and the initial conditions (\ref{eq:initial_condition_ak}). Then the acceleration of the rigid body motion of the disc is calculated by (\ref{eq:acceleration_h2}), which reads within the one-mode approximation,
\begin{equation}
h''(t)=\frac{F_{exp}(t)}{M+m_a}-\alpha_2\,a_1''(t), \quad \alpha_2=\frac{M_c}{M+m_a}W_1(0)+\gamma W_{10}.
\label{eq:one_mode_h2}
\end{equation}
\begin{figure}
\centering
\includegraphics[width=90mm,height= 78mm]{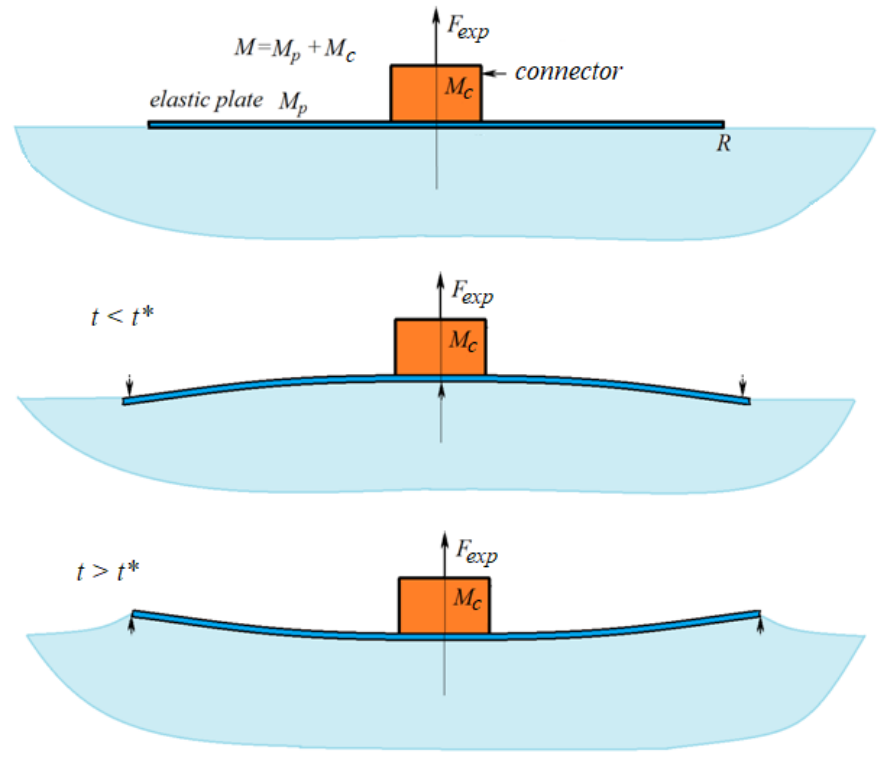}
\caption{\label{fig:sketch_motion}Sketch of the expected motion during the initial stages of the elastic disc lifting from the water surface.}
\end{figure}
The theoretical acceleration of the disc centre, $a(t)$, is calculated by equation (\ref{eq:theoretical_center_acceleration}),
\begin{equation}
a(t)=\frac{F_{exp}(t)}{M+m_a}+\alpha_3\,a_1''(t), \quad \alpha_3 = \frac{M_p+m_a}{M+m_a}W_1(0)-\gamma W_{10}.
\label{eq:acceleration_theoretical}
\end{equation}
For the conditions of our experiments, $\alpha_2=0.59$ and $\alpha_3=2.61$.
Therefore, the effect of the elastic deflection of the disc on the rigid motion of the disc is about four times smaller than on the acceleration of the disc centre.

\begin{figure}
\centering
\begin{tabular}{lr}
\hspace{-3mm}\includegraphics[width = 0.5\textwidth]{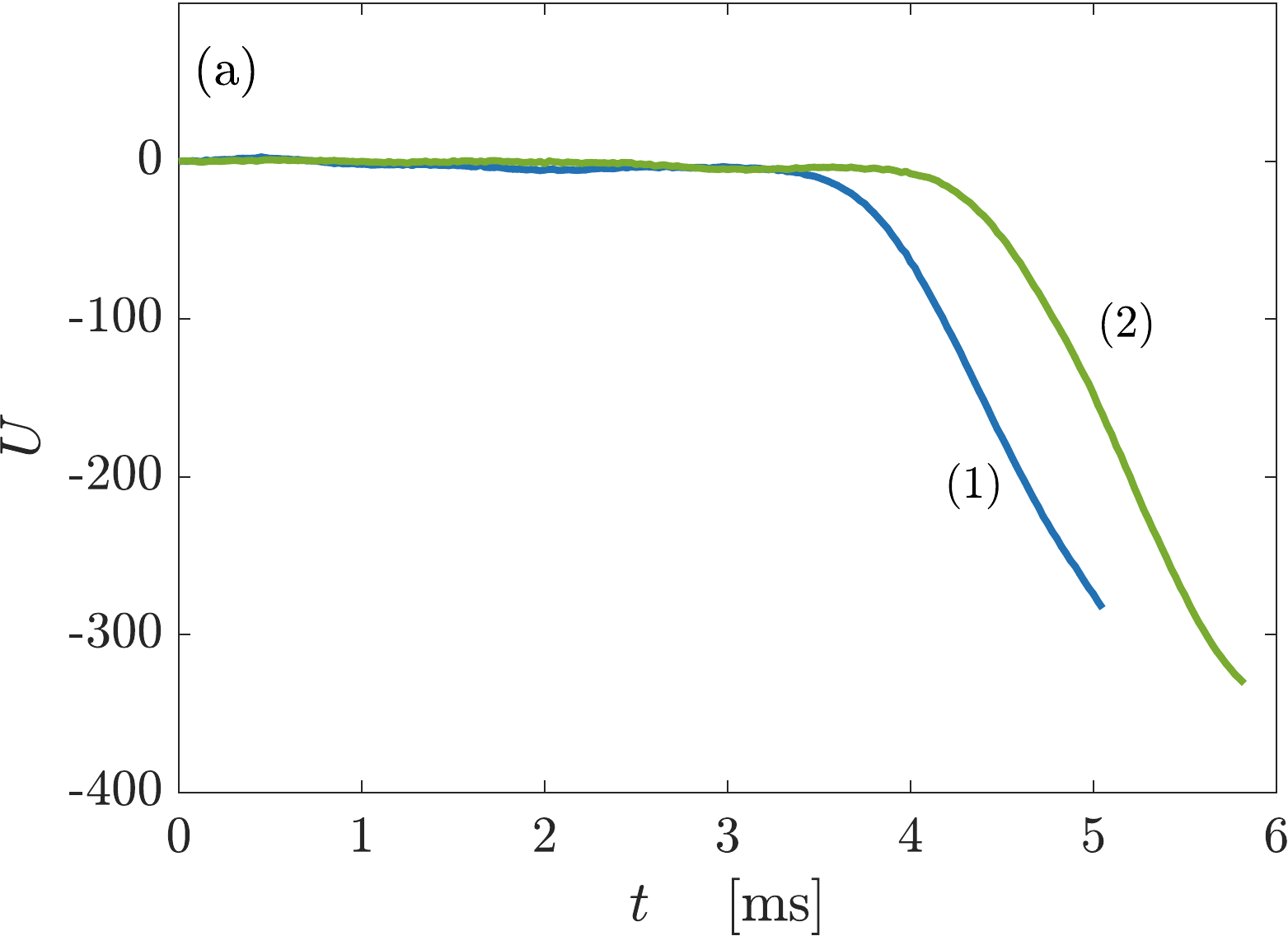} &
\includegraphics[width = 0.48\textwidth]{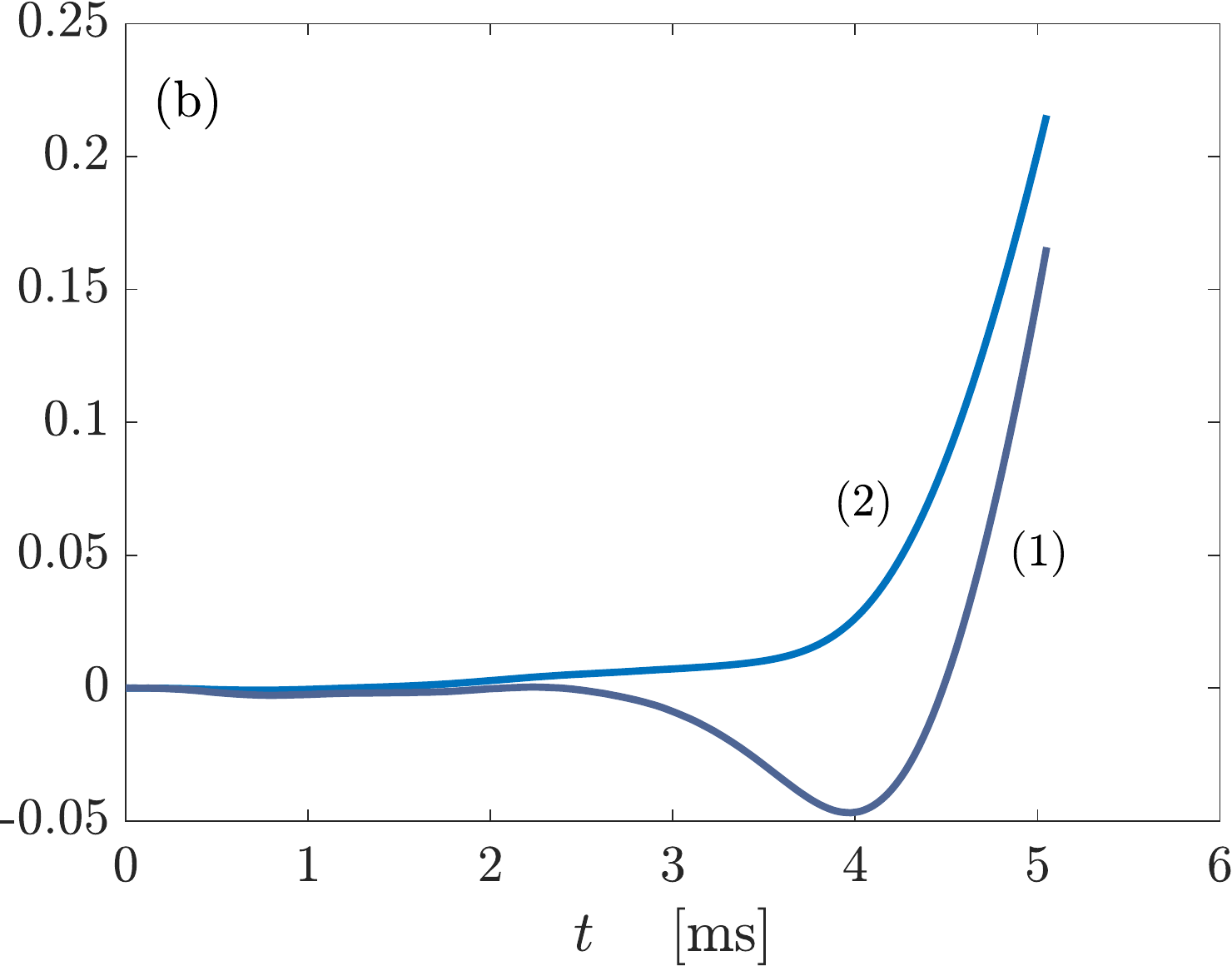}
\end{tabular}
\caption{\label{fig:pressure_coefficients}(a) Pressure coefficients $U(t)$ for two repetitions of session 3 (1 and 2). (b) $w_{t}(R,t)$ (line 1) and $V(t)$ (line 2) for session 3, repetition 1}.
\end{figure}

\begin{figure}
\centering
\begin{tabular}{lrlr}
\hspace{-3mm}\includegraphics[width=0.5\textwidth]{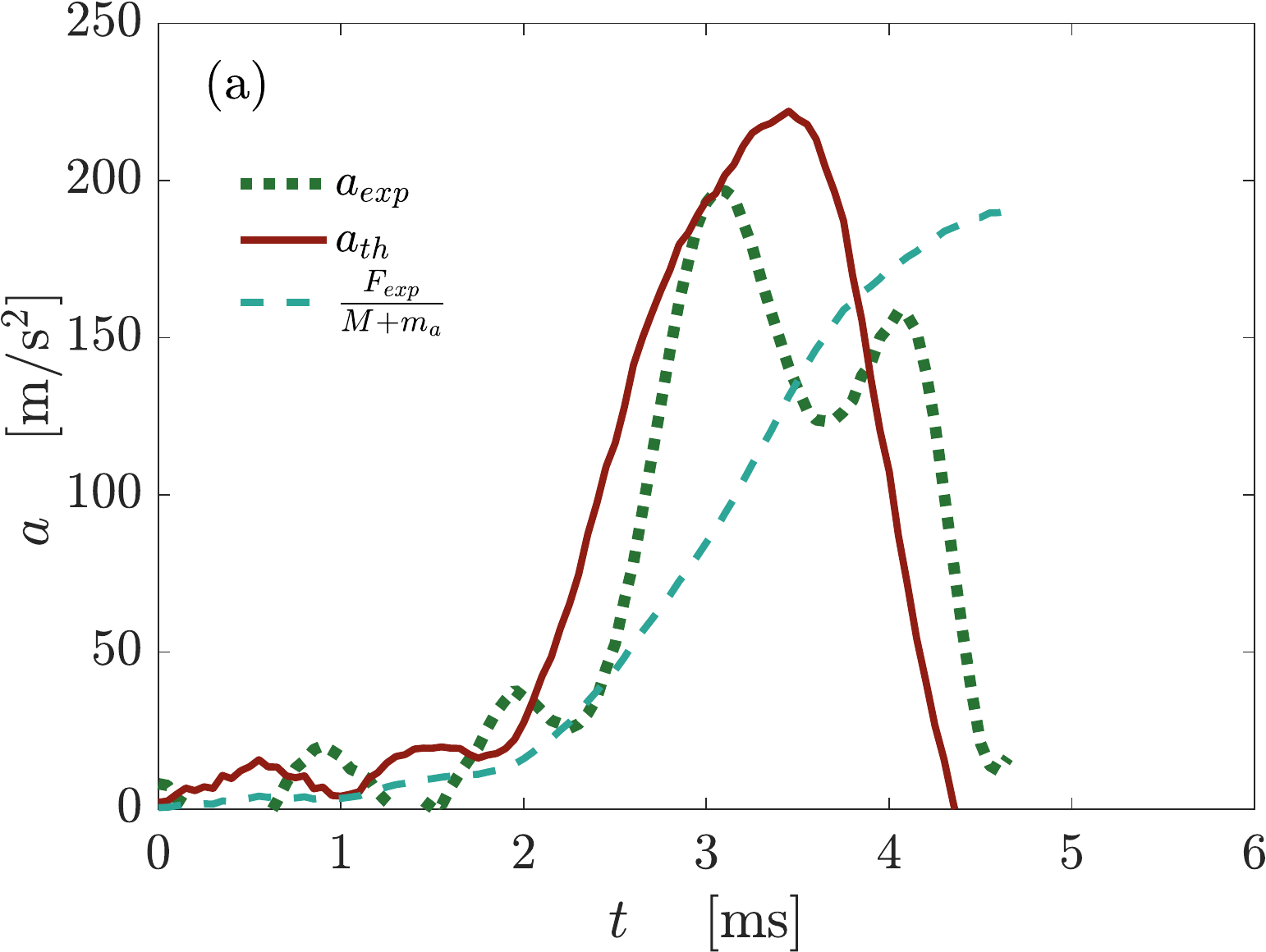} &
\includegraphics[width=0.5\textwidth]{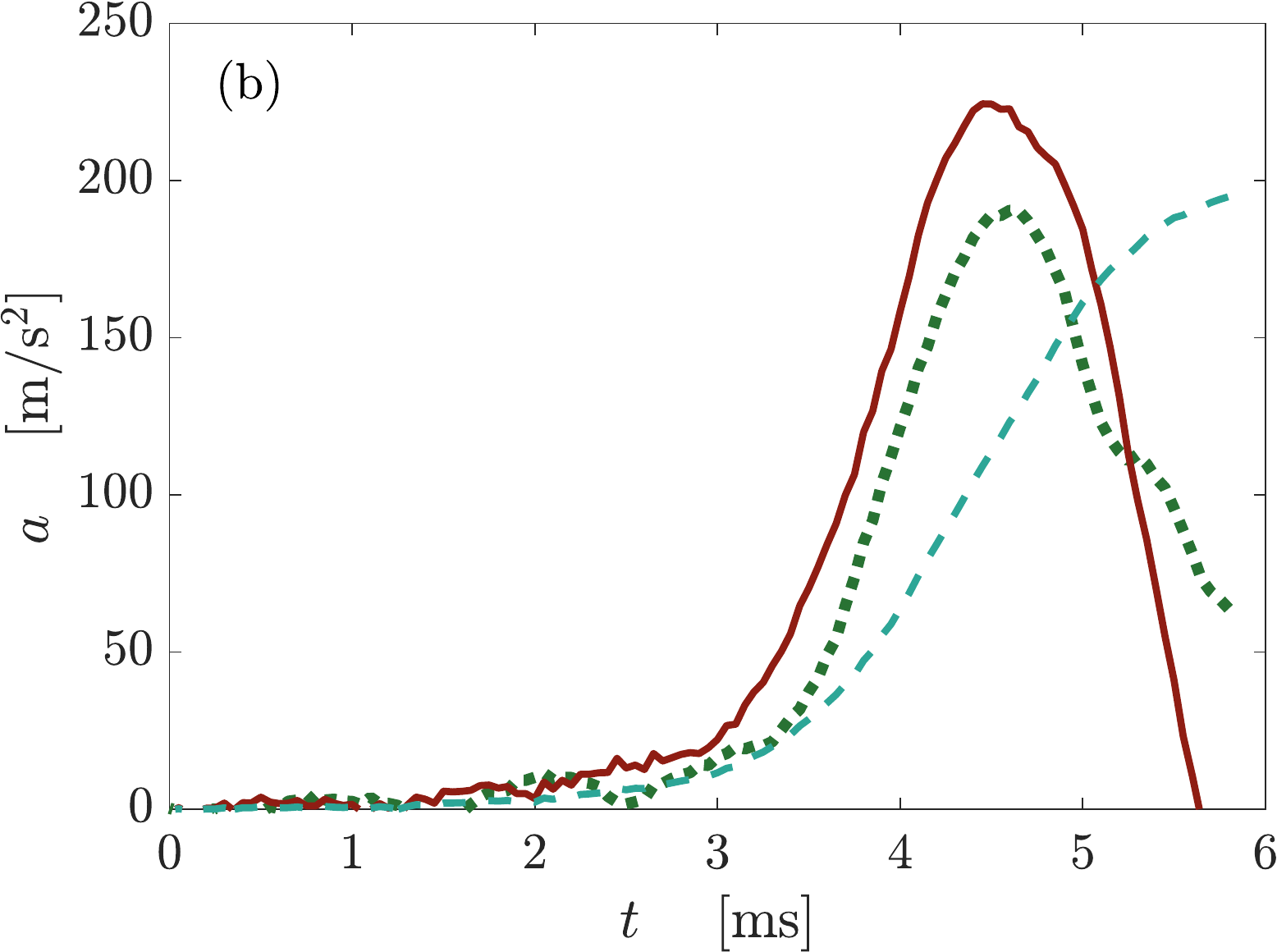}  \\
\hspace{-3mm}\includegraphics[width=0.5\textwidth]{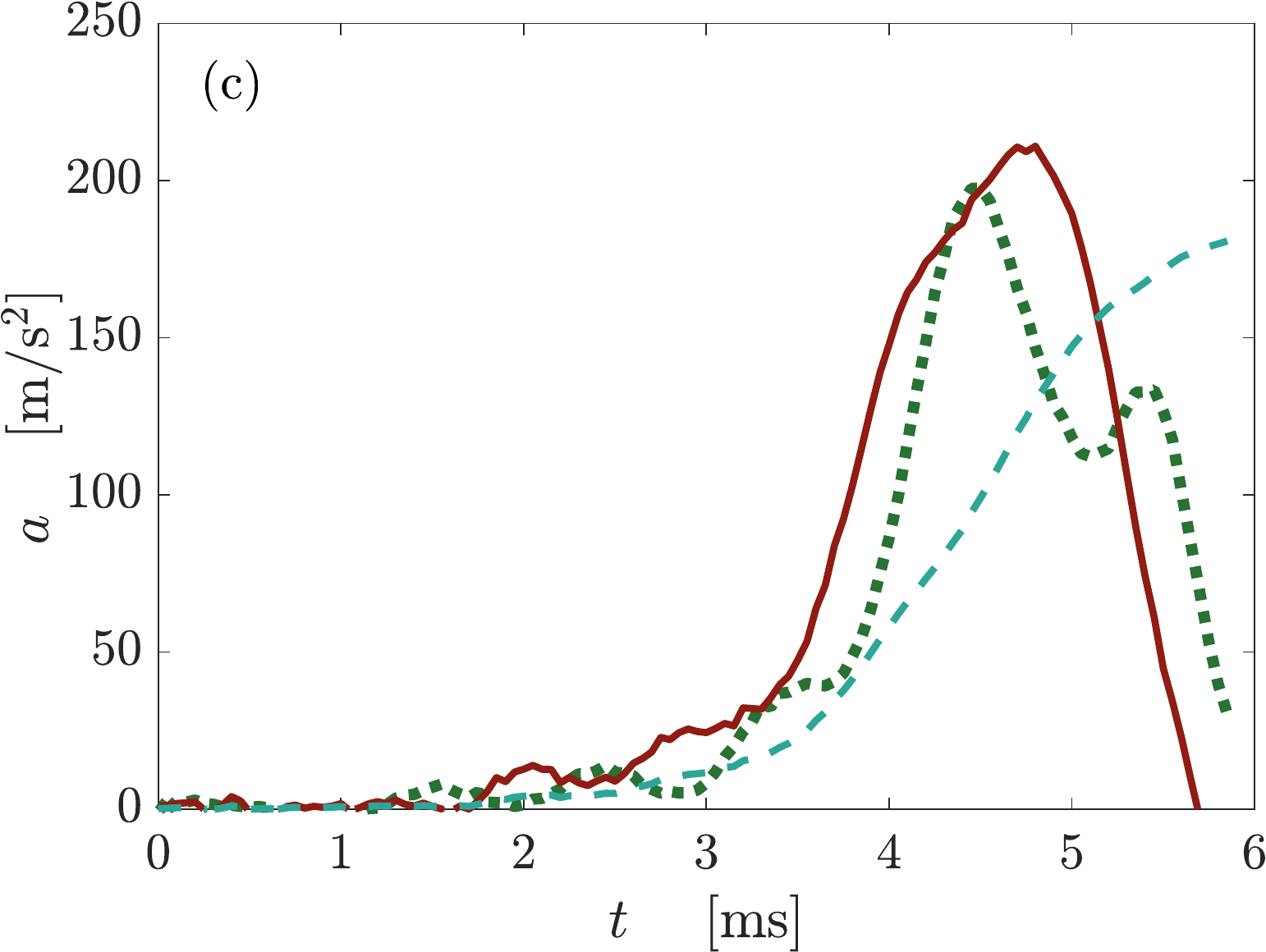} &
\includegraphics[width=0.5\textwidth]{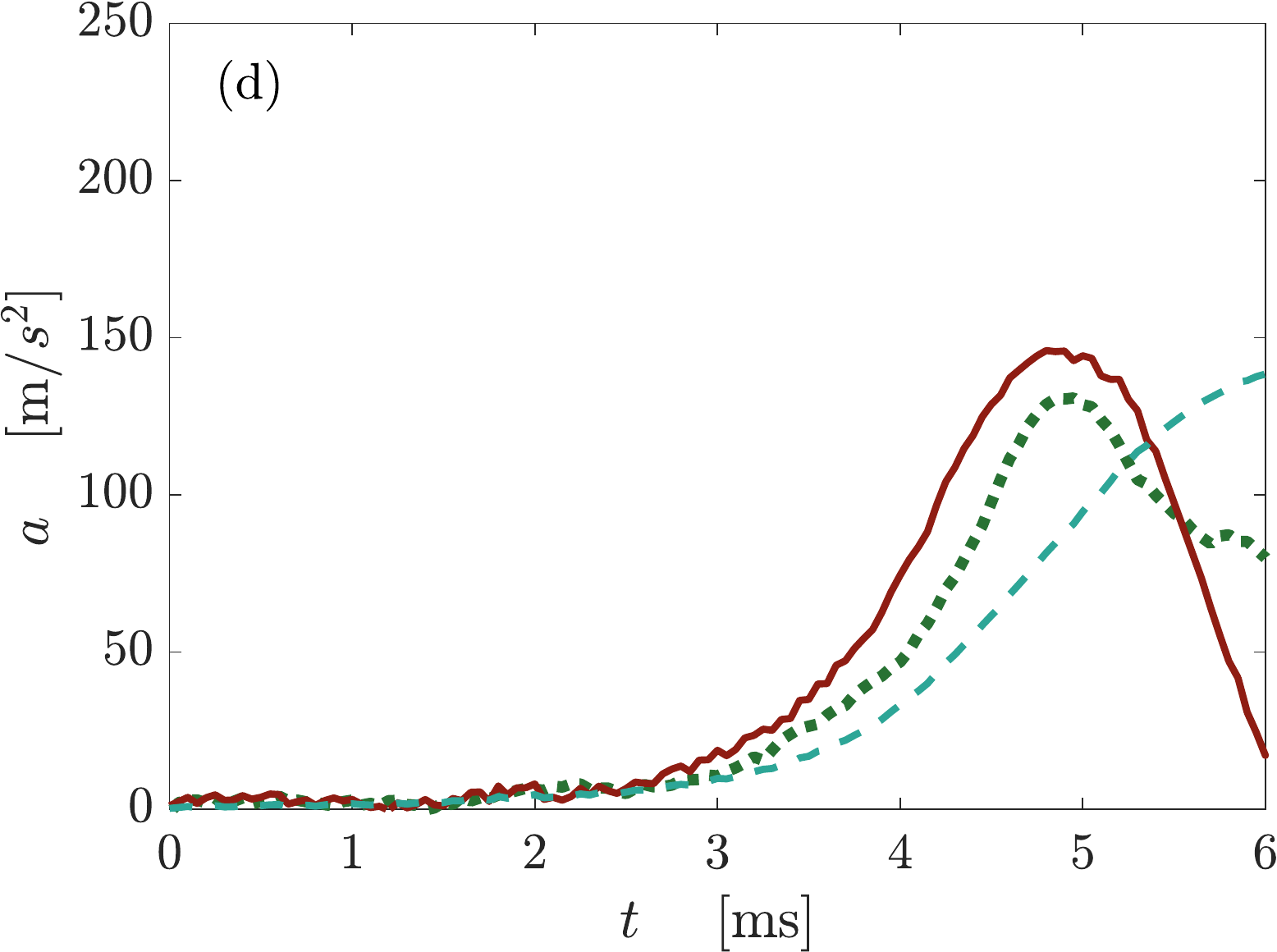}  
\end{tabular}
\caption{\label{fig:comparison_accelerations} Comparison of the theoretical accelerations at the disc centre with the experimental ones for four runs (session 3, repetitions 2, 3, 4, and 5). Dashed lines represent the normalised excitation force, $F_{exp}(t)/(M+m_a)$, solid lines the theoretical accelerations at the disc centre provided by equation (\ref{eq:theoretical_center_acceleration}) and the dotted lines are for the measured accelerations.}
\end{figure}

Figure \ref{fig:comparison_acceleration_components}a compares the measured acceleration of the disc centre (dotted line) with the theoretical predictions (equation \ref{eq:acceleration_theoretical}) (solid line)  for the first run of session 3. The first term in (\ref{eq:acceleration_theoretical}) is shown by the solid thin line. Therefore we can claim that accounting for the disc elasticity explains the relation between the measured force and acceleration.
More details about the disc motion are shown in figure \ref{fig:comparison_acceleration_components}b. It can be observed that the rigid-solid component of the acceleration $h''(t)$ (dash-dotted line) depends weakly on the elasticity of the disc and is close to the acceleration predicted by the rigid disc model (equation (\ref{eq:Fh_rigid}), dashed line. The total acceleration of the disc centre (solid line), $a(t)$, is made of two components within the one-mode approximation, $a(t)=h''(t)+a_1''(t)W_1(0)$, see equation (\ref{eq:theoretical_center_acceleration}). The elastic acceleration, $a_1''(t)W_1(0)$,  is shown by dotted line. It is larger than the rigid acceleration, $h''(t)$, for $0<t<4$ ms.
The acceleration of the disc edge, $w_{tt}(R,t)=h''(t) + a_1''(t)W_1(1)$, is shown by the solid line with dots markers. It is seen that this acceleration is negative for $0<t<3.8$ ms. Therefore, initially the edge of the disc goes down but the disc centre goes up. This situation is sketched in figure \ref{fig:sketch_motion}. The edge of the plate moves down when $0<t<t^*$, where $t^*$ is determined by the equation $w_t(R,t^*)=0$. The calculations provide $t^*=4.5$ ms for the case of figure \ref{fig:comparison_acceleration_components}. Initially the edge of the disc penetrates the water even though the main part of the disc exits the water. The radial velocity of the flow near the disc edge is given by equation (\ref{eq:radial_velocities}). Within the one-mode approximation, the vertical velocity of the disc edge,
\begin{equation}
w_t(R,t)=h'(t)+a_1'(t)W_1(1),
\label{eq:wt_at_edge}
\end{equation}
the coefficient $V(t)$ in (\ref{eq:radial_velocities}),
\begin{equation}
V(t)=h'(t)+a_1'(t)Q_1(1),
\label{eq:Vt_edge}
\end{equation}
and the coefficient,
\begin{equation}
U(t)=-h''(t)-a_1''(t)Q_1(1),
\label{eq:Ut_edge}
\end{equation}
in the asymptotic formula for the pressure,
\begin{equation}
P(r,0,t)\sim \frac{2\rho} \pi U(t)\sqrt{R^2-r^2}\qq (r\to R-0),
\label{eq:pressure_under_disc}
\end{equation}
near the contact line before it starts to move are depicted in figure \ref{fig:pressure_coefficients}.
Physically speaking the contact line cannot move if the pressure under the disc edge is greater than the atmospheric pressure, when $U(t)$ given by equation (\ref{eq:Ut_edge}) is positive, see equation (\ref{eq:pressure_under_disc}). The calculations provide that $U(t)$ is very close to zero for $0 < t < 4$ ms and quickly decreases after this time. The contact line does not move also if the radial velocity of the flow at the disc edge is positive, $V(t)$ given by equation (\ref{eq:Vt_edge}) is negative, see equation (\ref{eq:radial_velocities}). The calculations show that $V(t)$ is positive but very small for $0 < t < 3$ ms. The disc edge moves down for $0 < t < 4.6$ ms, see line 1. In figure \ref{fig:pressure_coefficients}b. Therefore figure \ref{fig:pressure_coefficients} shows that the generalised exit theory predicts that the contact line is unlikely to move before $t=3.5-4$ ms which is in good agreement with the experiments (see figure \ref{fig:wetted_radius}a).

The theoretical accelerations $a(t)$ calculated for different forcing functions $F_{exp}(t)$ in different experimental runs are compared with the measured accelerations in figure \ref{fig:comparison_accelerations}. These results confirm that taking into account the elastic deflection of the disc explains the non-monotonic relation between the measured forces and measured accelerations. Even though the the specific details of the measured forces and accelerations are rather sensitive to the precise initial conditions of each run, the theory predicts well the maximum accelerations and their duration.

\begin{figure}
\centering
\includegraphics[width = 0.53\textwidth]{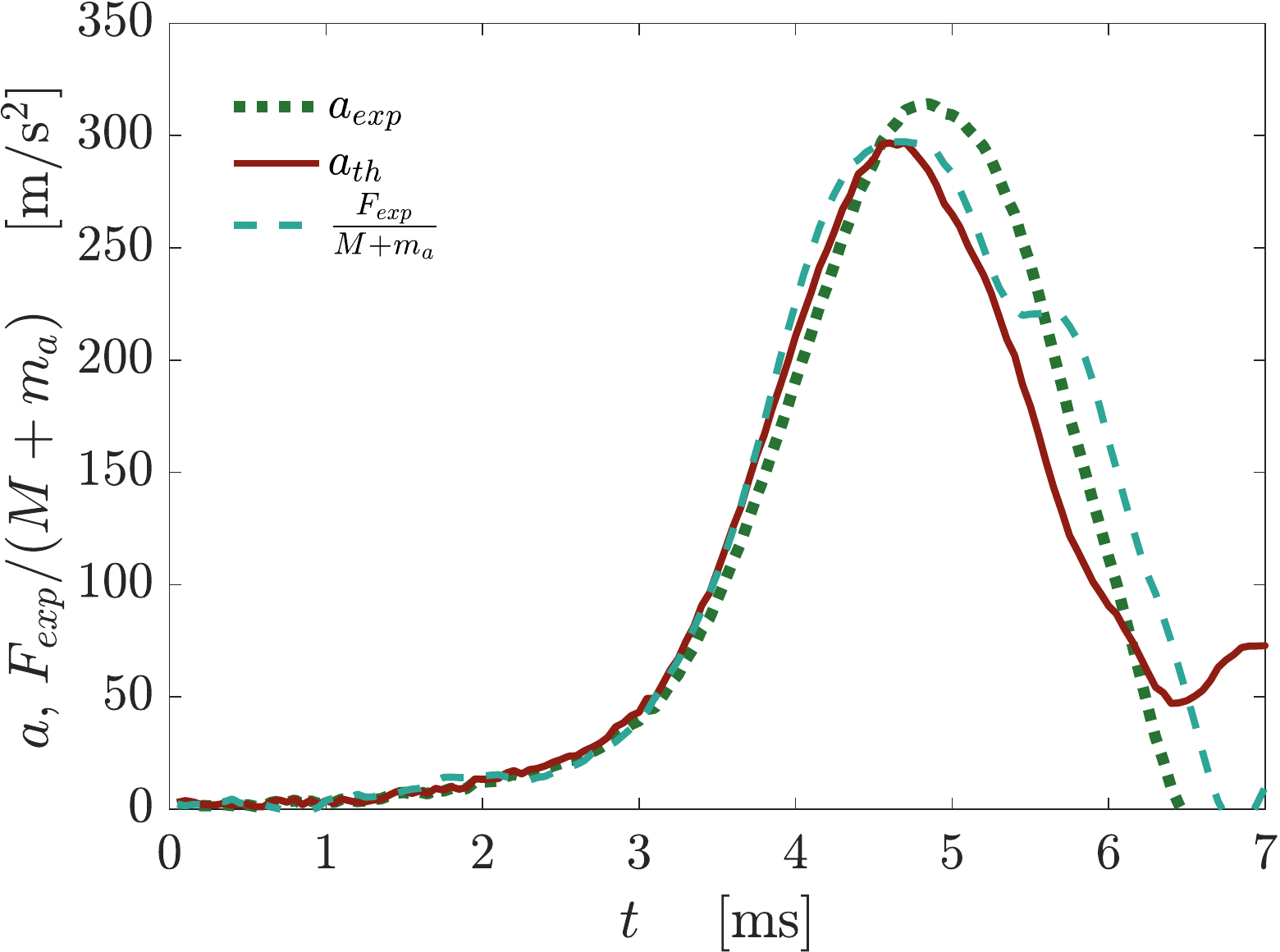}
\caption{\label{fig:dry_plate}Comparison of the normalised force, measured and theoretical accelerations during a dry test, session 1, where $m_a = 0$.}
\end{figure}

To prove that elastic effects are negligible in the absence of water, we show in figure \ref{fig:dry_plate} the experimental and theoretical accelerations for session 1. In this case the rigid disc model predicts well the results with $m_a = 0$.

\begin{figure}
\centering
\includegraphics[width = 0.48\textwidth]{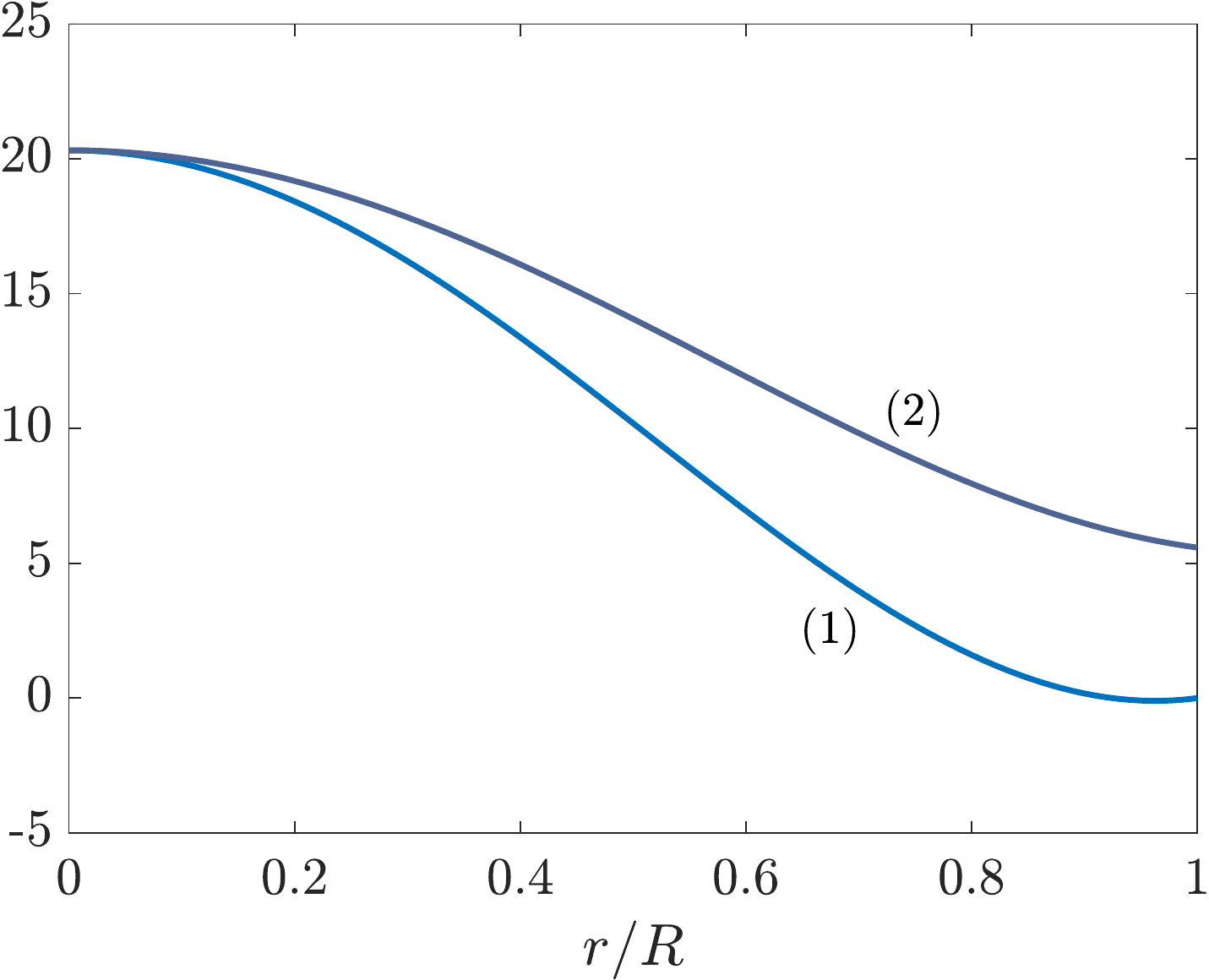}
\caption{\label{fig:stresses}Distributions of radial and tangential normalised stresses along the disc radius, curve (1) corresponds to $\tilde \sigma_r( \tilde{r})$ and (2) to $\tilde \sigma_t( \tilde{r})$.}
\end{figure}

To conclude this subsection, a few comments can be made about the distribution of stresses in the disc. The radial $\sigma_r$ and tangential $\sigma_t$ stresses on the disc surface are given by
\begin{equation}
\sigma_r(R \tilde{r}, t)=\frac{Eh}{2R^2(1-\nu^2)}\Big(W_1''(\tilde r)+\frac\nu{\tilde r} W_1'(\tilde r)\Big)a_1(t),
\label{eq:radial_stress}
\end{equation}
\begin{equation}
\sigma_t(R \tilde{r}, t)=\frac{Eh}{2R^2(1-\nu^2)}\Big(\nu W_1''(\tilde r)+\frac 1{\tilde r} W_1'(\tilde r)\Big)a_1(t),\qq (\tilde r<1),
\label{eq:tangential_stress}
\end{equation}
in the one-mode approximation. For the conditions of figure \ref{fig:comparison_acceleration_components}, the theoretical amplitude of the elastic mode $a_1(t)$ increases monotonically from zero to $4\times 10^{-4}$ mm at the end of the measurements. The functions  $\tilde \sigma_r( \tilde{r})=W''_1(\tilde r)+\nu W'_1(\tilde r)/\tilde r$ and $\tilde \sigma_t( \tilde{r})=\nu W''_1(\tilde r)+ W'_1(\tilde r)/\tilde r$ are shown in figure \ref{fig:stresses}. Their maximum values are achieved at $\tilde r= 0$ and are equal to each other. For the acrylic plate used in the experiments, the tensile strength is 69 MPa and the factor $Eh/[2R^2(1-\nu^2)]$ in (\ref{eq:radial_stress}) and (\ref{eq:tangential_stress}) is approximately 1.5 GPa. Therefore the disc can be fractured starting from the centre if the amplitude of the elastic mode reaches $a_1 = 2.3$ mm.
Thus, since this displacement is large enough to be observed in the videos, unlike the ones we have, we conclude that we are far away from plate failure conditions.

\subsection{Effect of plate thickness}

\begin{figure}
\centering
\begin{tabular}{lrlr}
\hspace{-3mm}\includegraphics[width=0.5\textwidth]{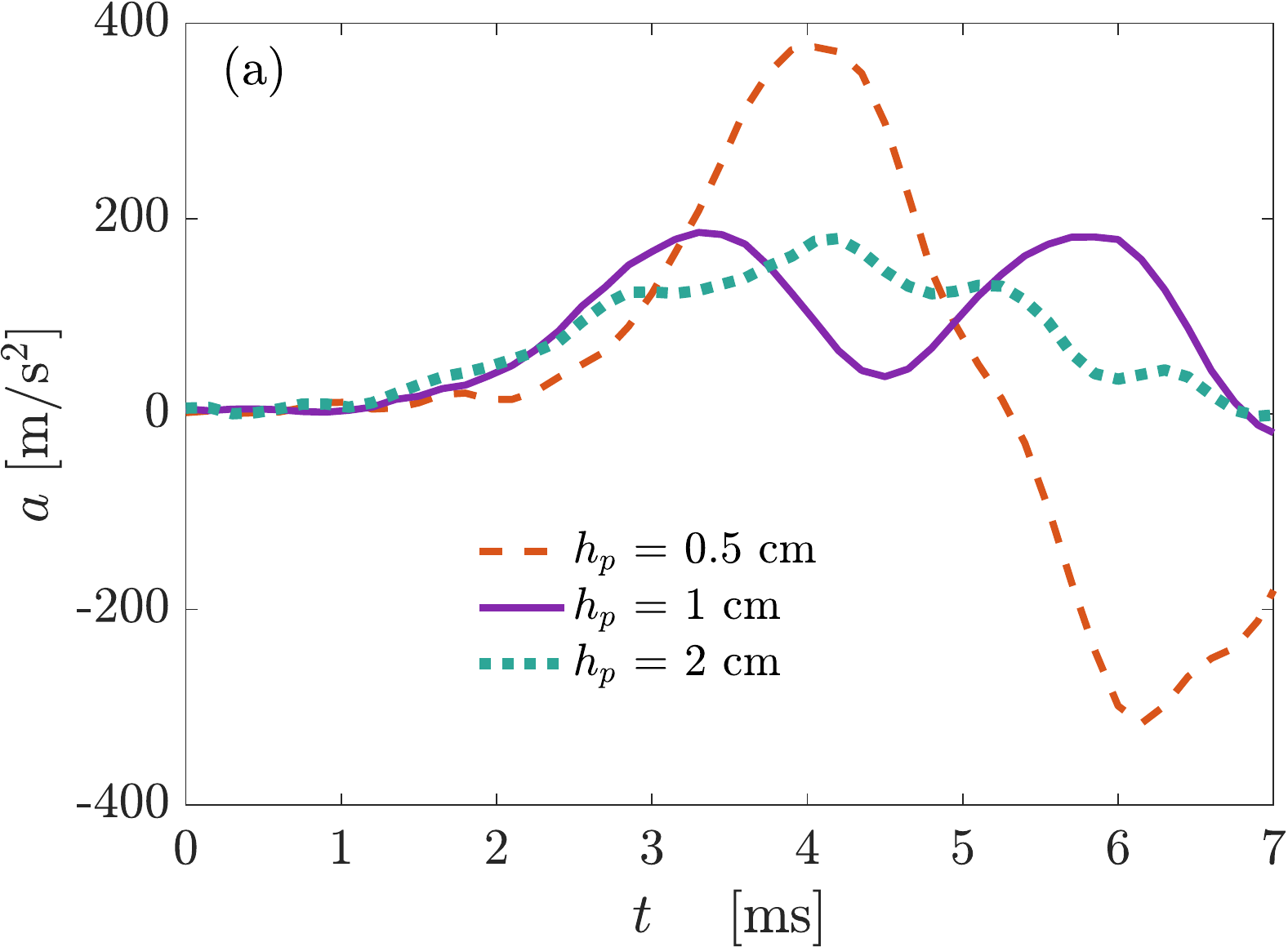} &
\includegraphics[width=0.5\textwidth]{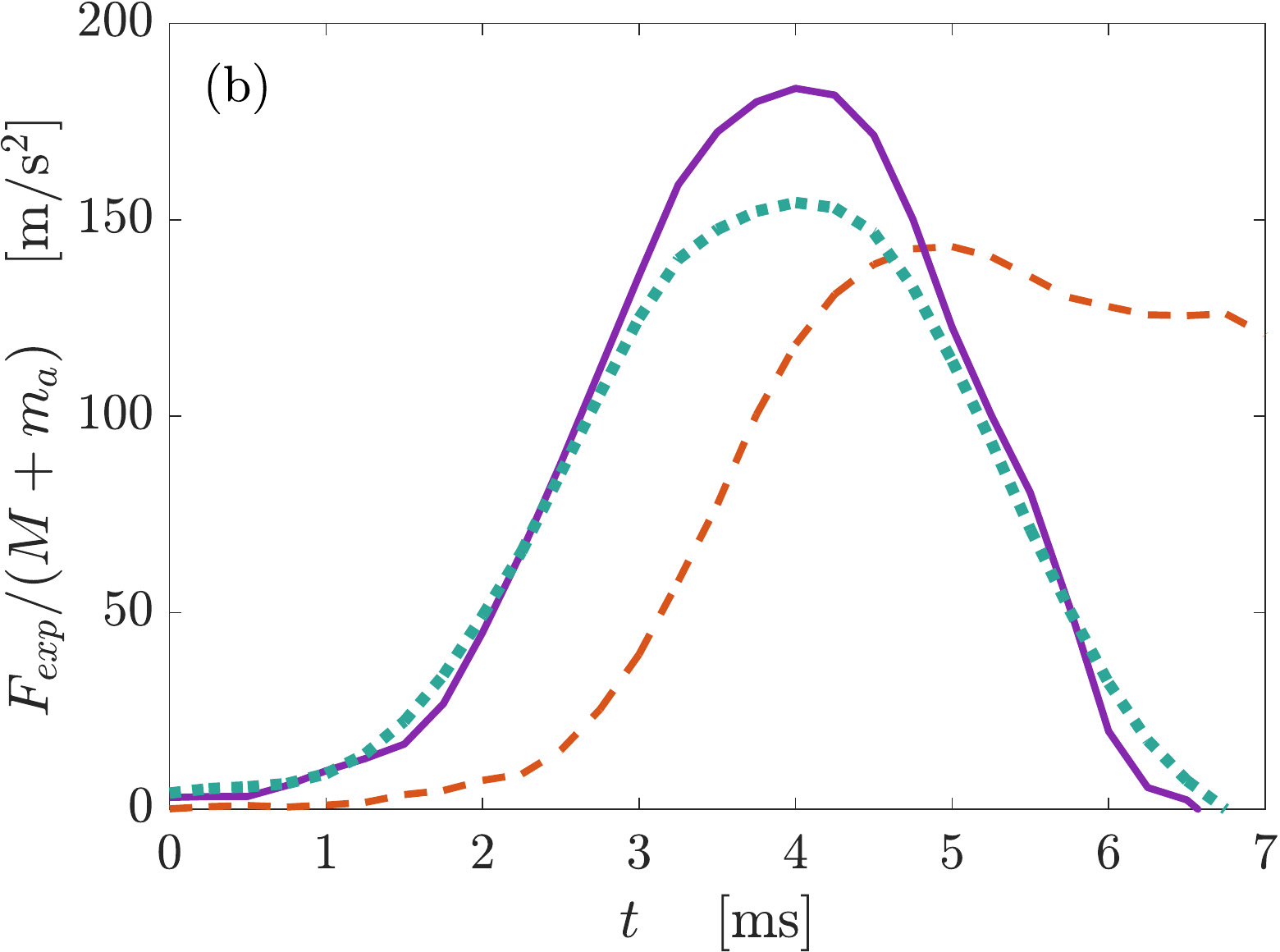}  \\
\hspace{-3mm}\includegraphics[width=0.5\textwidth]{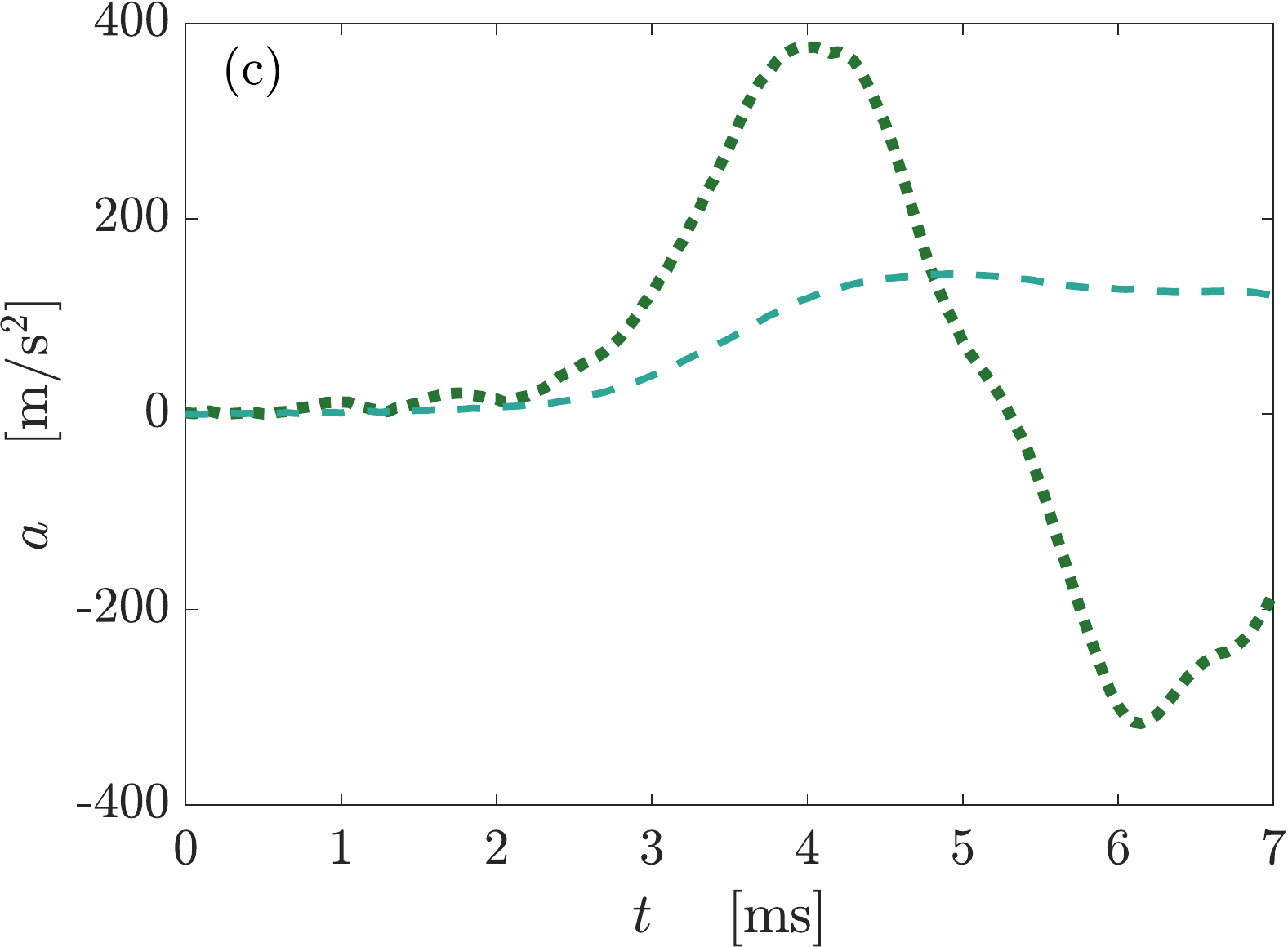} &
\includegraphics[width=0.5\textwidth]{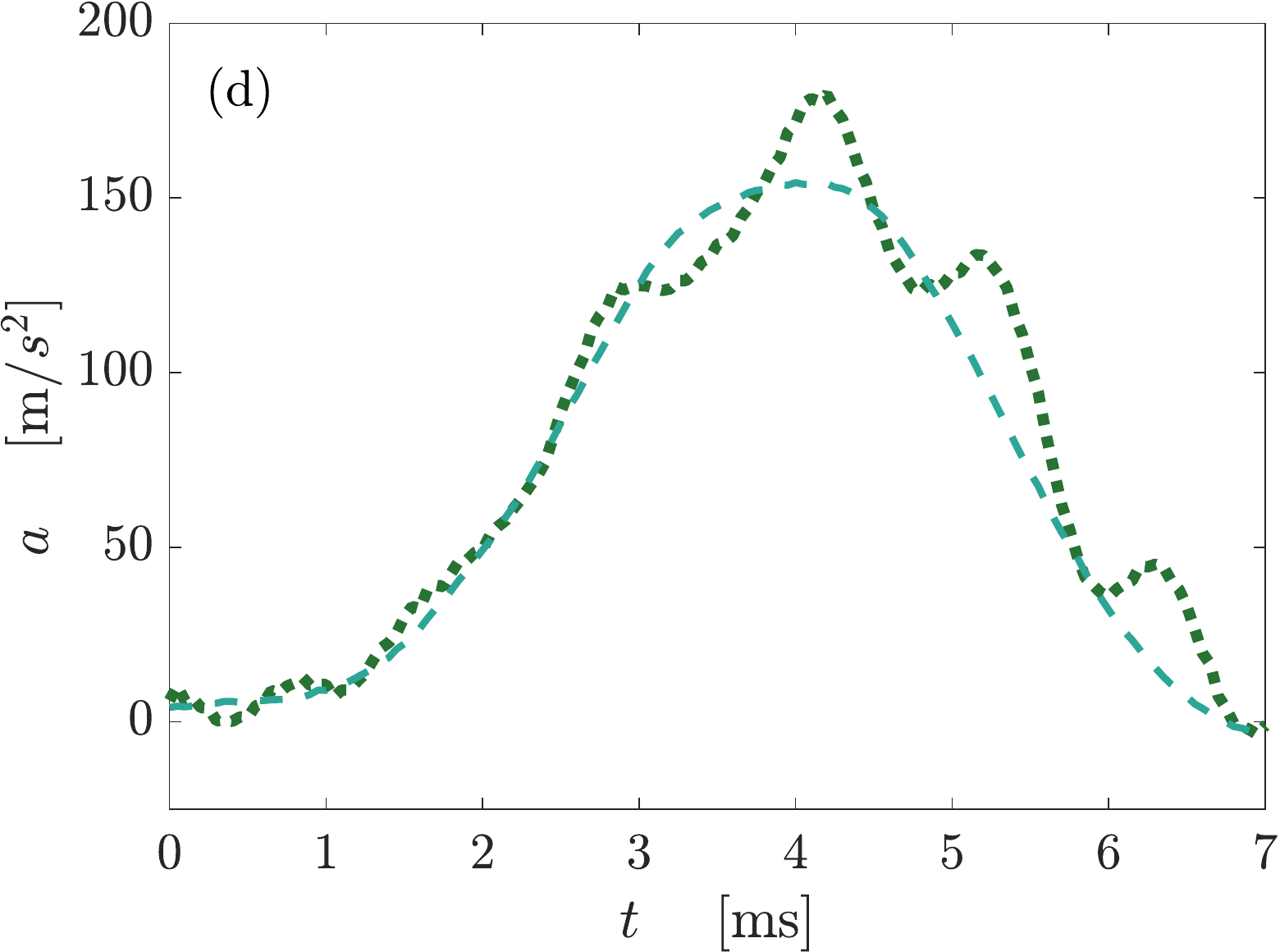}  
\end{tabular}
\caption{\label{fig:effect_plate_thickness}Comparison of experimental accelerations (a) at the disc centre for the normalised measured forces (b) for three different plate thicknesses, 0.5, 1, 2 cm respectively. Comparison of experimental acceleration (dotted lines) and normalised measured forces (dashed lines) for (c) the plate thinkness, $h_p$, 0.5 cm and (d) 2 cm.}
\end{figure}

To further support the conclusion that elastic effects are essential during the initial stages of the plate exit, we provide here additional experiments with two discs of thicknesses $h_p = 0.5$ cm and $h_p = 2$ cm, see table \ref{tab:properties_discs}. As expected, the vibrations of the thicker disc have smaller amplitude and higher frequency than those of the thinner one, as depicted in figure \ref{fig:effect_plate_thickness}. In general, the period of these oscillations is reasonably well captured by the theory. It should be stressed that the results presented here have been obtained using the measured properties of the plates and that, once these parameters are fixed, the theory has no adjustable parameters and its only input is the excitation force measured experimentally.
\begin{table}
    \centering
    \begin{tabular}{ccccc}
    $h_p$ (cm) & $M_p$ (kg) & $M$ (kg) & $\omega_{wet,1}$ (s$^{-1}$) & $T_{wet,1}$ (ms) \\ \hline
    0.5 & 0.216 & 0.414 & 775 & 8.10 \\
    1 & 0.432 & 0.630 & 1945 & 3.23 \\
    2 & 0.864 & 1.07 & 5034 & 1.25
    \end{tabular}
    \caption{\label{tab:properties_discs}Properties of the three different discs used in experiments.}
\end{table}
\begin{figure}
\centering
\begin{tabular}{lr}
\hspace{-3mm}\includegraphics[width=0.5\textwidth]{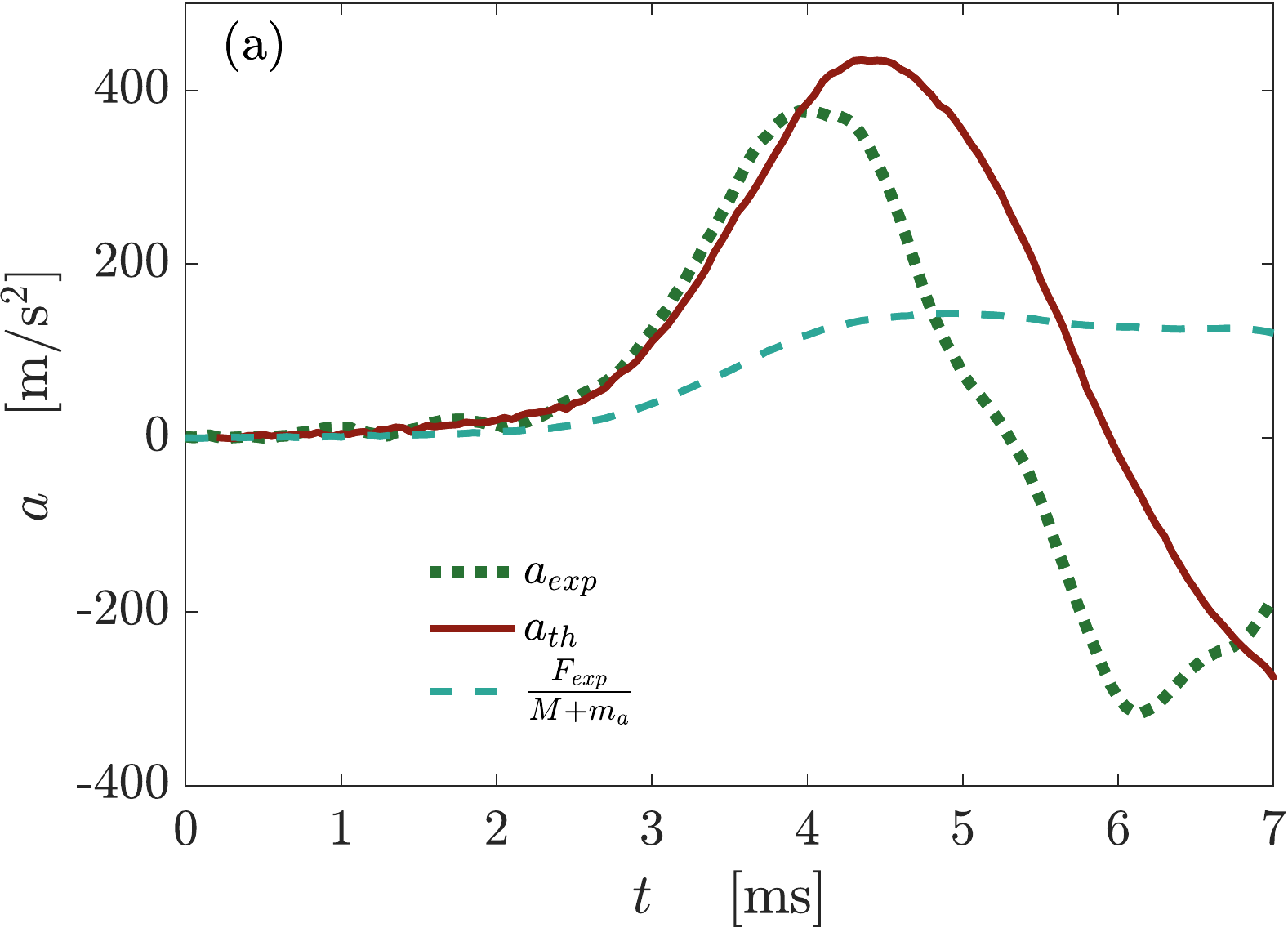} &
\includegraphics[width=0.5\textwidth]{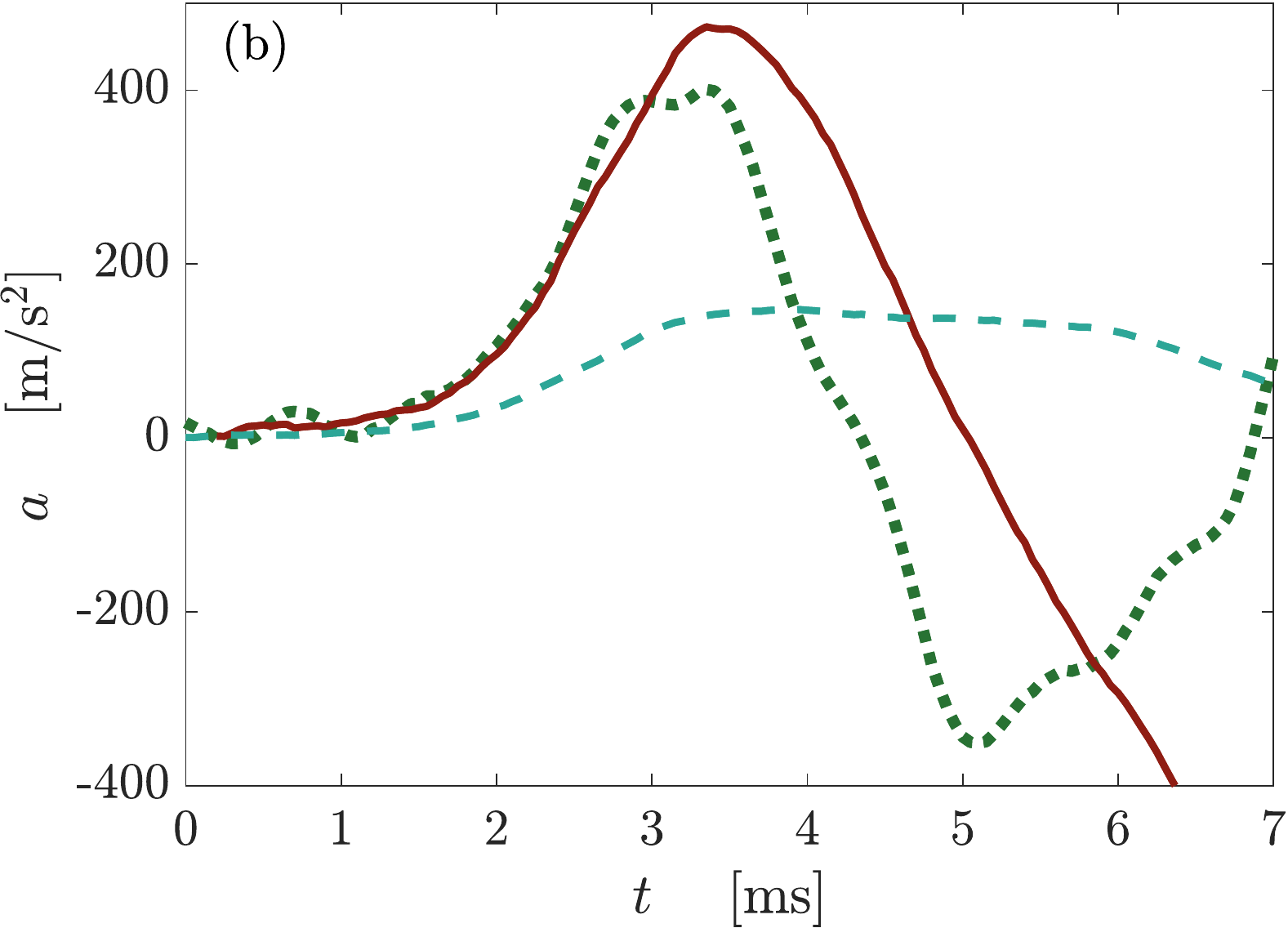}  \\
\hspace{-3mm}\includegraphics[width=0.5\textwidth]{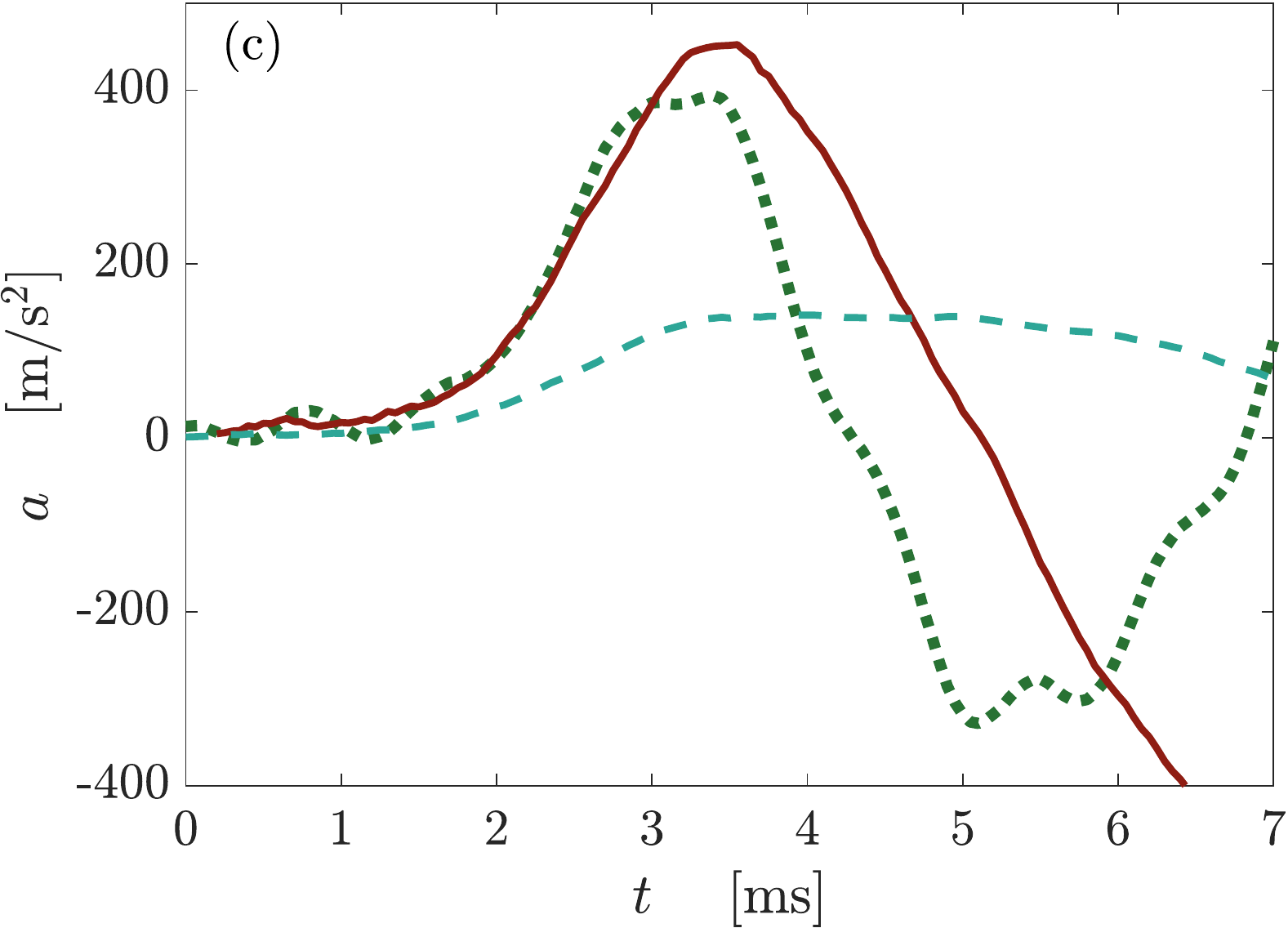} &
\includegraphics[width=0.5\textwidth]{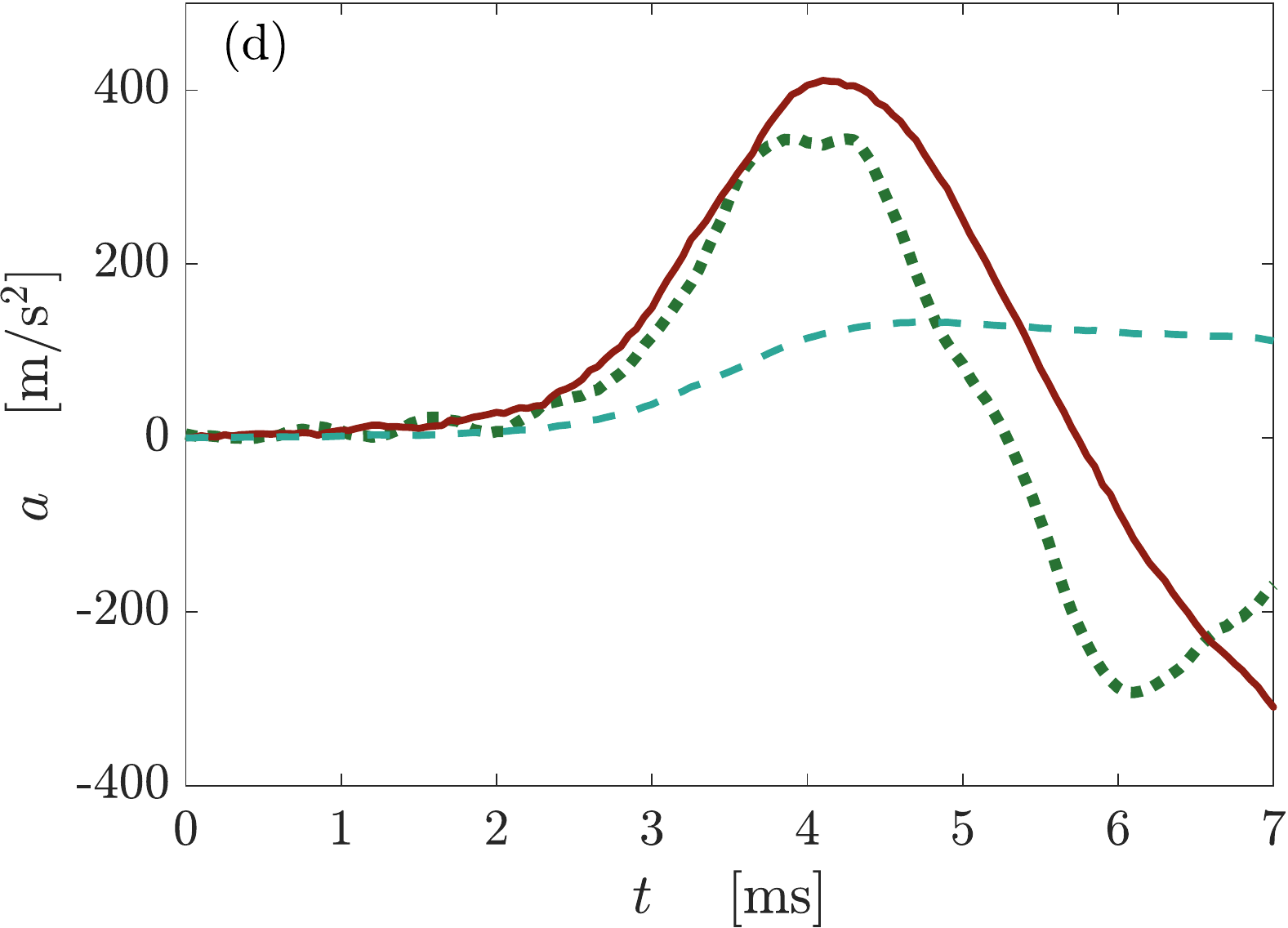}  
\end{tabular}
\caption{\label{fig:runs_05} Comparison of theoretical accelerations at the disc centre with the experimental ones (session 5) in four runs for disc thinckness, $h_p = 0.5$ cm. Dashed lines are for the normalised excitation force, $F_{exp}(t)/(M+m_a)$, solid lines are for the theoretical accelerations at the disc centre provided by equation (\ref{eq:theoretical_center_acceleration}) and dotted lines are for the measured acceleration.}
\end{figure}

For the thinner plate, the first peak of the acceleration is very well described by the theory, as illustrated in figure \ref{fig:runs_05}, albeit the second one exhibits an amplitude substantially smaller than the calculated one. We hypothesize that this disagreement is caused by several effects. First, it should be reminded that, in the theory, we assume that the force is applied at the center, whereas in reality it is applied on a non-axisymmetric region. Second, non-linearity leads to the appearance of higher-order modes, including non-axisymmetric ones, which are excited more easily in a thinner plate, since their --longer-- natural period is closer to the duration of the experiment. A third explanation may be sought in the fact that our theory does not account for any source of dissipation. Due to the smaller elastic energy stored in a thin plate, the effects of dissipation are expected to be more important in relative terms. Finally, because of its smaller mass, a thin plate leaves the water surface at a slightly larger acceleration than a heavier one, thus the assumption that the virtual mass is that evaluated at $t=0$ becomes questionable.

\begin{figure}
\centering
\begin{tabular}{lr}
\hspace{-3mm}\includegraphics[width=0.5\textwidth]{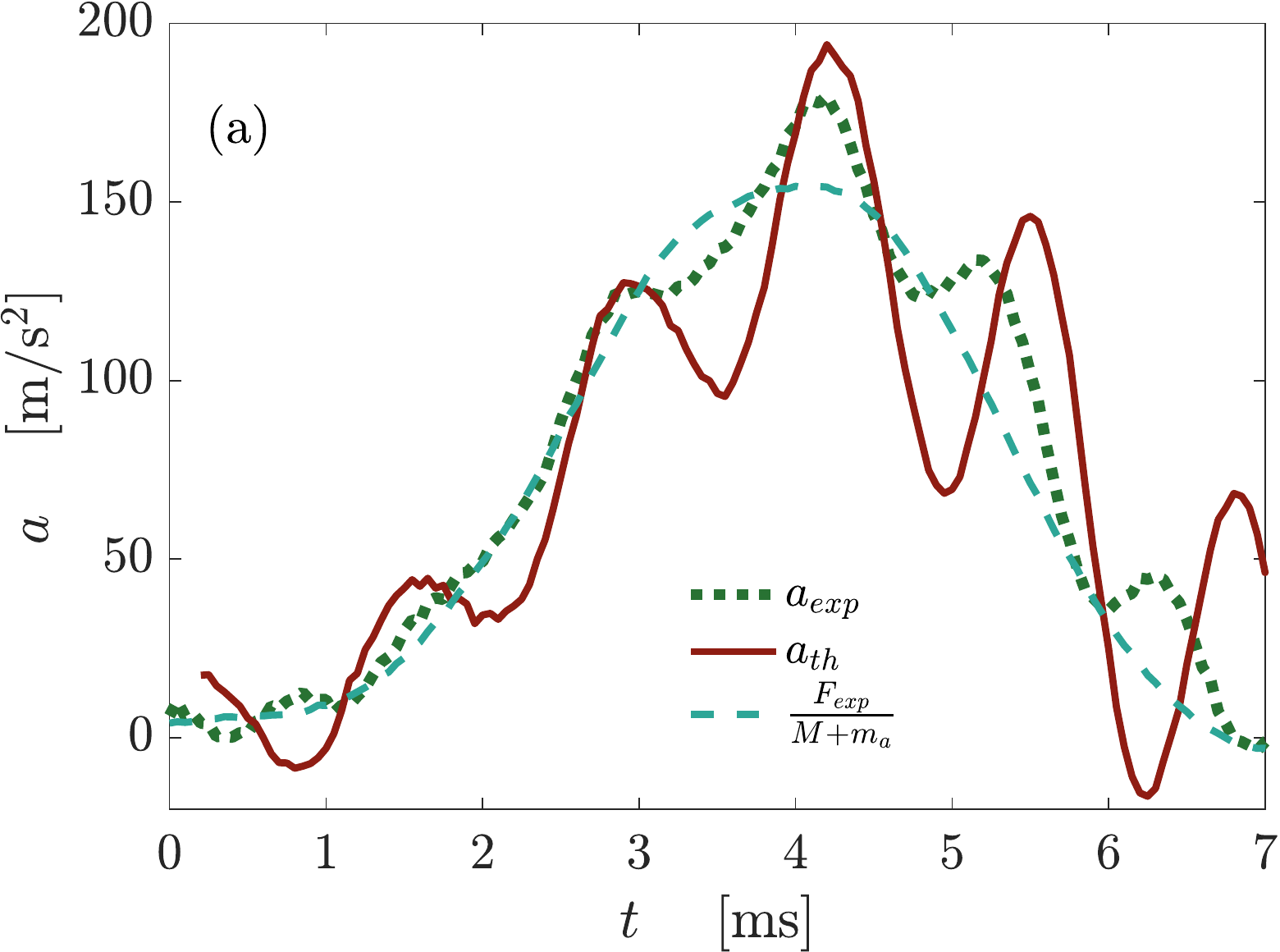} &
\includegraphics[width=0.5\textwidth]{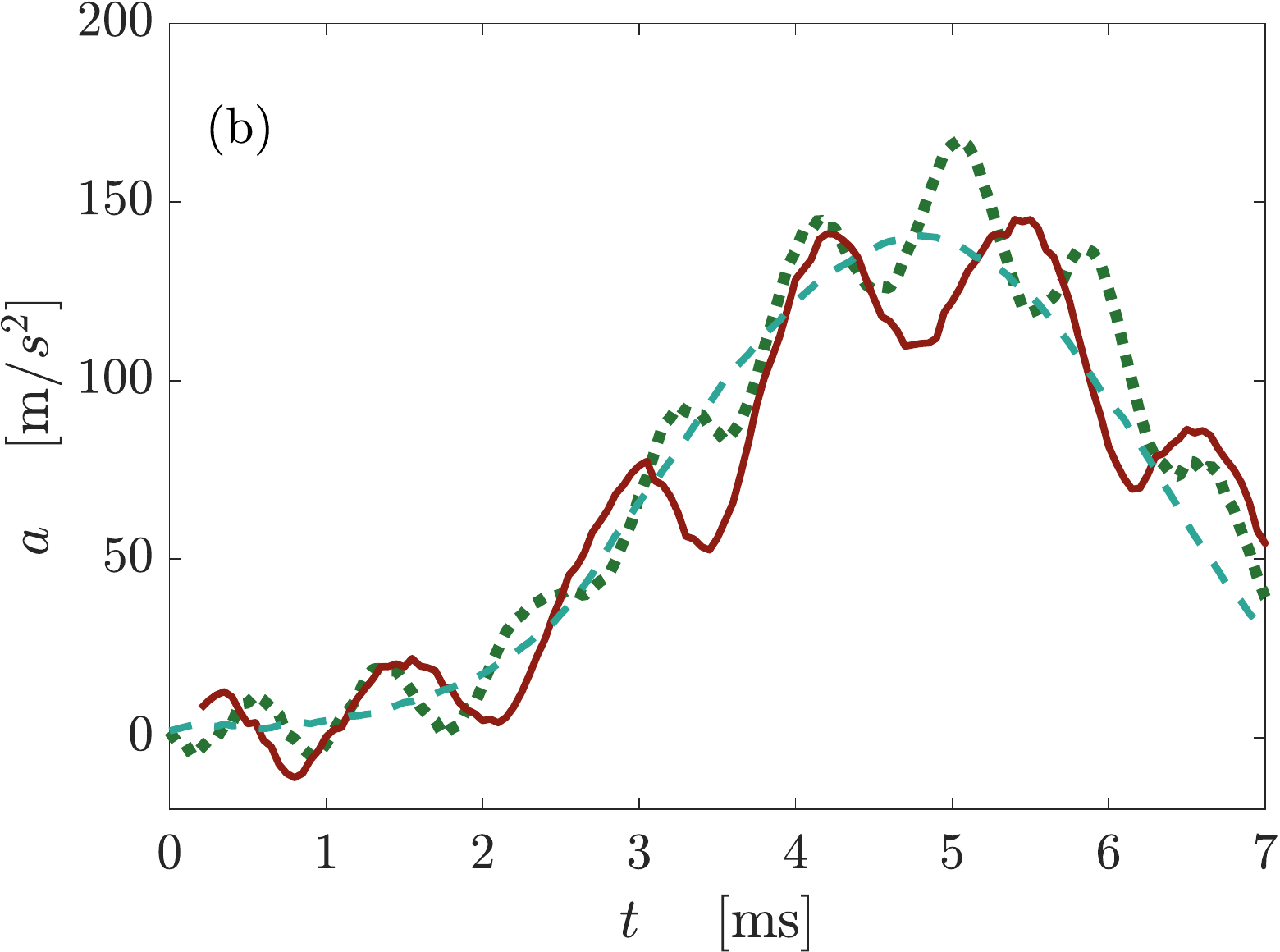}  \\
\hspace{-3mm}\includegraphics[width=0.5\textwidth]{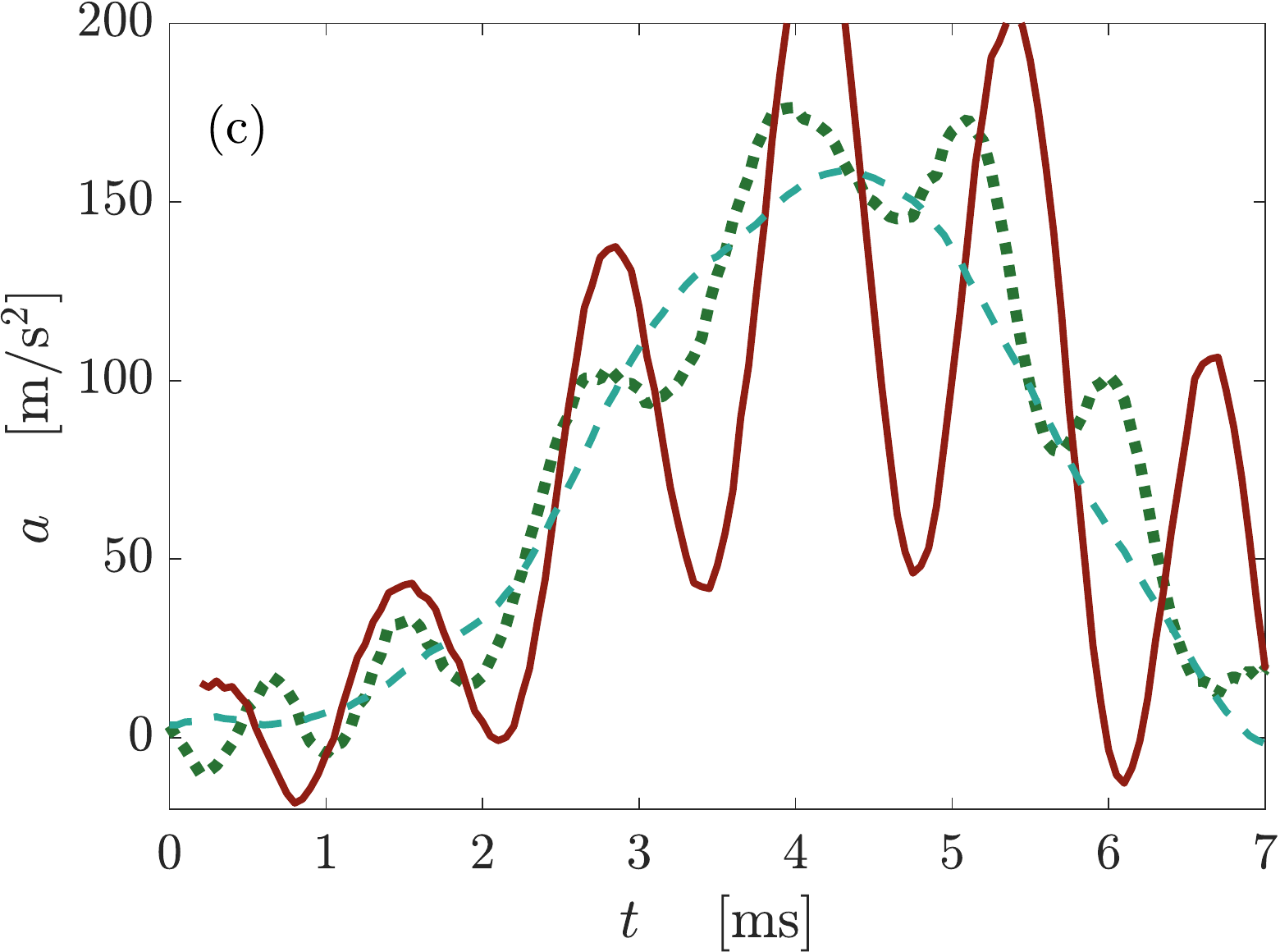} &
\includegraphics[width=0.5\textwidth]{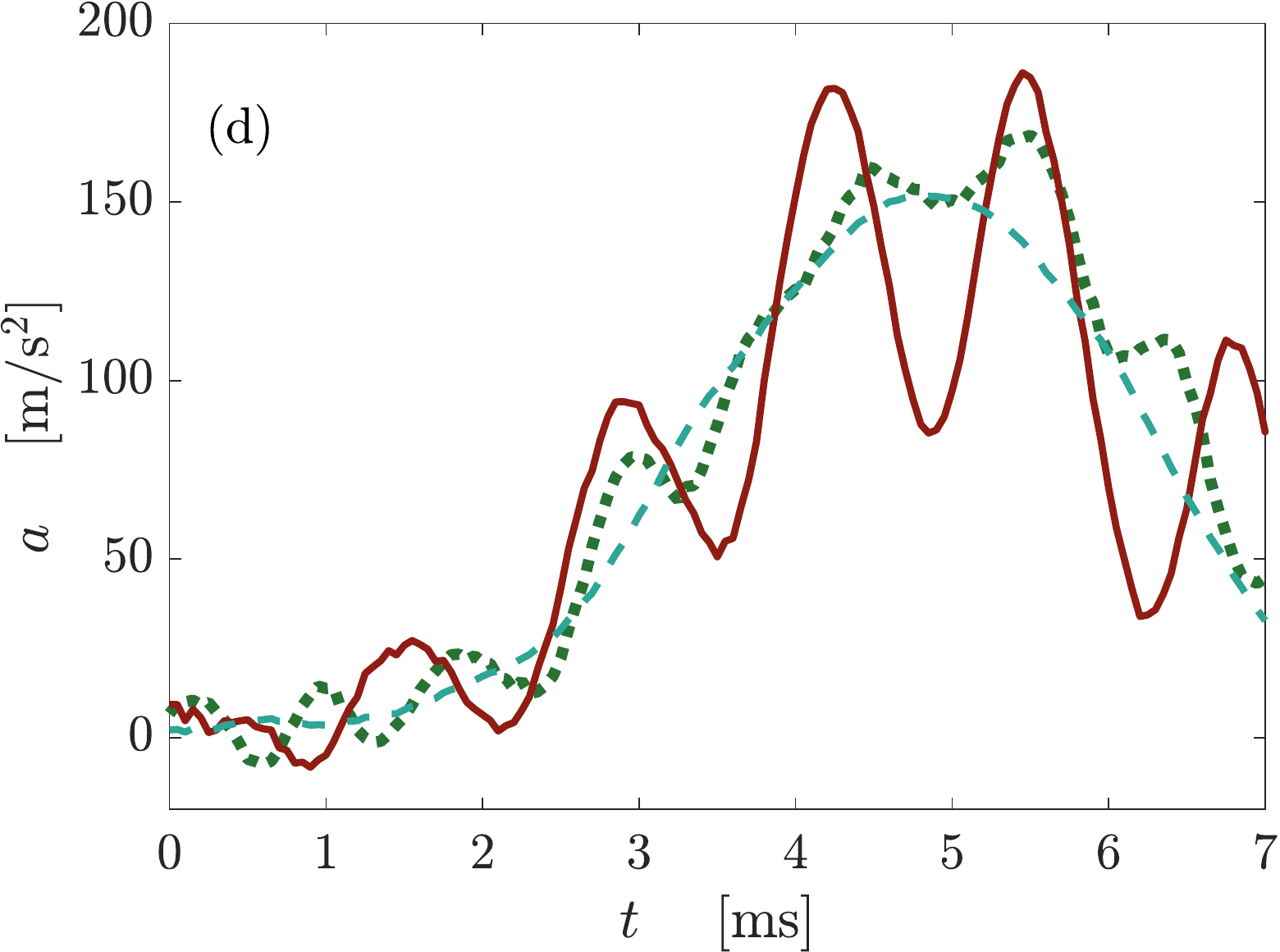}  
\end{tabular}
\caption{\label{fig:runs_20}Comparison of theoretical accelerations at the disc centre with the experimental ones (session 4) in other four runs for disc thinckness, $h_p=2$ cm. See figure \ref{fig:runs_05} for an explanation of the different curves.}
\end{figure}

Regarding the results for the thick plate, $h_p = 2$ cm (see figure \ref{fig:runs_20}), the theory predicts accelerations which oscillate around the experimental ones, albeit with a larger amplitude, which we attribute again to the lack of dissipation in the model. Despite this discrepancy, the period these oscillations is well predicted by the theory. Thus, we find fair to state that the theory predicts better the experimental behavior in this case. Indeed, for a thick plate we expect the arguments used in the previous paragraph for a thin plate to reverse, which explains the better agreement between theory and experiments.

In view of these results, we find reasonable to claim that the generalised linear theory of water exit presented here can be successfully used to describe the motions of an elastic body that leaves the water surface at a large acceleration, such that its elastic vibrational modes are excited.

\section{Conclusions}

We designed, built and tested a facility to investigate the sudden lifting of a plate from a water surface. The results of the experimental campaign carried out in this facility for a circular plate, a disc, have been presented and analysed using the added mass theory. Several non-intuitive phenomena have been observed in the experiments. First the acceleration of the plate close to the place where the external force is applied does not monotonically increase in time even if the external force does. Second, the wetted part of the plate does not start instantly to decrease with the plate lifting. Instead, there is an initial interval of time during which the wetted part of the plate does not shrink. It was found that both these phenomena are governed by the elastic properties of the plate and the interaction between the liquid flow and the plate deflection.

It was inferred from these observations that the modelling of the suction force experienced by a rigid plate that leaves a water surface at a large acceleration is complex, due to the interplay between the body motion and the free surface dynamics. Nonetheless, at the very early stages of the motion this modelling was greatly simplified by linearizing the equations and boundary conditions through exploiting the fact that the displacements of the body and the water surface are very small. In particular, this allowed us to impose the boundary conditions at the undisturbed (flat and horizontal) free surface and initial body position. The linearised exit theory of rigid bodies \citep{Korobkin2013} predicts that the body acceleration is proportional to the driving force, where the proportionality coefficient depends on the added mass of the body.

Although at very short times this approach yields reasonable results, as was shown for instance by the numerical simulations of \cite{Korobkin2017} as well as by our experimental observations (see figure \ref{fig:dry_wet}b), eventually the acceleration predicted by the theory significantly departs from the acceleration at the center of the plate obtained experimentally. Most notably, the observed accelerations even decrease in response to a monotonically increasing force.

This disagreement has motivated the development of a generalised linear theory of water exit that takes into account the elastic response of the plate to large accelerations. This new theory is based on the same ideas of \cite{Korobkin2013} but includes, for the first time, the deformation dynamics of the plate through the linear elastic theory of thin plates.

Taking advantage of the linear nature of both the hydrodynamic and elastic problems, the disc displacement was expressed as a series of normal elastic modes of a free-free circular disc supported at its center. The theoretical results obtained with only one mode agree fairly well with experiments, while keeping the complexity of the solution at a reasonable level. We stress here again that this theory has no free parameters to adjust, and all the required inputs are determined experimentally.

Besides yielding accelerations close to the experimental ones, the hydroelastic theory of water exit predicts that the contact line between the liquid, the air and the plate will remain attached initially to the edge during several milliseconds, which is also in excellent agreement with our observations using high-speed movies. It is also interesting to point out here that, once the contact line detaches from the plate edge, it recoils as the plate instantaneous height raised at the power of 2/3, as predicted by the self-similar solution obtained by \cite{KorKhaba2017} for the flow close to the corner of a plate leaving a water surface at a large acceleration.

We would like to highlight the surprising result that plates like the ones used in our experiments, seemly very stiff, exhibit elastic effects strong enough to completely alter their surface exit dynamics. We have shown that the additional local inertia caused by the liquid motion is responsible for this elastic behavior. This effect was not observed in those experiments where the plate does not touch the water, see figure \ref{fig:dry_plate}. In these cases, the acceleration and the applied force nearly follow each other within the experimental error margin.

Despite the, in general, good agreement with the experiments, our theoretical model still exhibits some discrepancies, specially for small plate thicknesses. We attribute those to: (a) the non-axisymmetric way in which the force is applied to the disc, (b) the appearance of higher-order and non-asisymmetric modes, not considered in our calculations, (c) the absence of a dissipation mechanism that surely exists in the experiments, and (d) the assumption that the disc surface remains completely wet at all times.

Even taking into account the assumptions just mentioned, we believe that it is fair to conclude that the hydroelastic theory presented here emerges as a simple yet useful tool to compute the fluid-structure interaction in water exit phenomena occurring at large accelerations.

P.V.M. and J.R.R. acknowledge the work of the Spanish Ministry of Economy and Competitiveness through grants DPI2014-59292-C3-1-P, DPI2015-71901-REDT, and DPI2017-88201-C3-3-R, partly funded by Europan funds. T.I.Kh. and A.A.K. are supported by the NICOP research grant ``Vertical penetration of an object through broken ice and floating ice plate'' N62909-17-1-2128, through Dr. Salahuddin Ahmed.

\bibliographystyle{jfm}

\end{document}